\documentclass[10pt]{article} 

\PassOptionsToPackage{sort}{natbib}
\usepackage[preprint]{tmlr}

\usepackage{graphicx}
\usepackage{url}            
\usepackage{booktabs}       
\usepackage{amsfonts}       
\usepackage{nicefrac}       
\usepackage{microtype}      
\usepackage[dvipsnames]{xcolor}         
\usepackage[colorlinks=true,allcolors=NavyBlue]{hyperref}       
\usepackage{amsmath}
\usepackage{doi}

\DeclareMathOperator{\ReLU}{ReLU}

\DeclareMathOperator{\sign}{sign}

\DeclareMathOperator{\tr}{tr}
\DeclareMathOperator{\linspan}{span}

\title{Stimulus symmetries can confound \texorpdfstring{\\}{} representational similarity analyses}

\author{\name Farhad Pashakhanloo \email fpasha@fas.harvard.edu \\
      \addr Center for Brain Science\\
      Harvard University
      \AND
      \name Jacob A. Zavatone-Veth \email jzavatoneveth@fas.harvard.edu\\
      \addr Center for Brain Science and Society of Fellows\\
      Harvard University}

\def\month{MM}  
\def\year{YYYY} 
\def\openreview{\url{https://openreview.net/forum?id=XXXX}} 

\begin{document}
\maketitle
\begin{abstract}
  What can representational similarity matrices (RSMs) tell us about a neural code? As the popularity of these summary statistics grows, so too does the need for a more complete characterization of their properties. Here, we show that symmetries in network inputs can confound RSM-based analyses. Stimulus symmetries render many representations functionally equivalent, but these different configurations can lead to different RSMs. These different RSMs reflect qualitatively different representational geometries, ranging from disentangled to maximally-mixed codes. We show that stochastic gradient descent or energetic regularization can generate sparse, drifting codes, leading in turn to drifting RSMs. Moreover, we demonstrate that these phenomena are present in networks trained to encode image data, where the symmetry is latent. Our results illustrate the challenges inherent in comparing nonlinear neural codes, when functionally-equivalent representations are not related by a simple rotation. 
\end{abstract}

\addtocontents{toc}{\protect\setcounter{tocdepth}{-1}}

\section{Introduction}

The sensory world is replete with symmetries. Most famously, the identity of a visual object is invariant to its position and pose \citep{rao1998learning}. To robustly interpret the external world, the brain must grapple with these symmetries as it encodes and learns from natural data: the functionality of a representation should not be affected by globally transforming the stimulus space according to the symmetry.

In machine learning, recent years have seen significant interest in designing neural network architectures that are constrained to respect stimulus symmetries \citep{bronstein2021geometric}. In particular, \textit{equivariant} neural networks are engineered such that their internal representations are linear representations in the sense of group theory. That is, if one transforms the input according to the symmetry, the encodings at a given hidden layer of the stimulus before and after the transformation are related by a global linear transformation, most simply a rotation \citep{bronstein2021geometric,cohen2015transformation,kondor2018convolution}. This property is advantageous mathematically, as it allows for decorrelation and disentangling as different symmetries act on the input \citep{cohen2015transformation,higgins2018disentangled}. Moreover, it makes it easy to understand when two representations are related by the action of the stimulus symmetry. However, it is a strong constraint which is not inherently satisfied by most neural networks, whether natural or artificial. 

How, then, should one analyze and understand neural representations of symmetric data? Perhaps most simply, given two representations---whether from different networks or from the same network at different stages of learning---how should one compare them? For both machine learning and for neuroscience, the search for principled measures of representational similarity is key to attempts to assess questions of convergence in encoding strategies, whether between artificial networks, between biological networks, or between brains and machines \citep{sucholutsky2025getting,huh2024platonic}.

Efforts to systematize this search have thus far focused mostly on designing metrics that are invariant to internal symmetries intrinsic to network architectures \citep{sucholutsky2025getting,godfrey2022symmetries,harvey2024what}. The motivation here is simple: two networks which are related by an intrinsic symmetry may erroneously appear different when viewed through the lens of a summary statistic that is not invariant to that symmetry. For example, one can compensate for global rotations of the encoding at one layer of a deep network by rotating the synaptic weights of the next layer. Thus, rotation-invariance is a reasonable desideratum for a summary statistic of deep representations. This invariance is realized by the matrix of dot-product similarities of encodings of a set of stimuli, known as the representational similarity matrix (RSM). Both as a precursor to downstream analyses---like representational similarity analysis (RSA) or centered kernel alignment (CKA)---or as an endpoint in itself, the RSM has gained broad popularity as a tool for understanding neural population codes \citep{kriegeskorte2008representational,kriegeskorte2021geometry,sucholutsky2025getting,harvey2024what,williams2024equivalence,han2023identification,onoo2026readout}. 

In contrast to the case of intrinsic symmetries, less is understood about how extrinsic symmetries of the data distribution affect representational comparisons outside the setting of explicitly equivariant network architectures. Here, we ask whether two representations related by a stimulus symmetry can have distinct RSMs, even though the symmetry leaves the underlying computational function unchanged. We refer to such gauge dependence as a \textit{confound}, in the statistical sense of a factor that can systematically alter the outcome of an analysis, and thus must be taken into account when that outcome is interpreted. We do not mean that gauge-dependence is intrinsically undesirable; depending on the scientific goal, it may be either a bug or a feature. Though previous works have demonstrated the existence of function-preserving transformations that change the RSM, the transformations they consider are far less restrictive than transforming the stimulus space according to a symmetry \citep{bo2025evaluating,ding2021grounding,davari2023reliability,braun2025not,ciernik2025objective}. Thus, how this highly-rigid class of representational transformations affect downstream similarity analyses remains unclear.

We show that stimulus symmetries generate families of functionally equivalent representations that are not, in general, related by an orthogonal transformation in representation space, and therefore have distinct RSMs. This highlights a possible mismatch between function-preserving stimulus symmetries and the orthogonal transformations to which RSMs are invariant. As a tractable model setting, we focus on manifold-tiling neural codes, in which the neurons' receptive fields form a regular lattice over stimulus space. The brain uses manifold-tiling codes to represent a range of physical and abstract variables, and they emerge through learning in a variety of model settings \citep{hubel1962receptive,sengupta2018tiling,gardner2022toroidal}. From a theoretical perspective, manifold-tiling representations are a simple example of a nonlinear neural code. For these codes, we show that the stimulus symmetry does not in general act as a rotation on the representation space, meaning that examining the RSM would lead one to believe that codes related by the symmetry are distinct. We illustrate this through a variety of controlled experiments, spanning synthetic and image data. Together, our results show how biologically-plausible nonlinear neural codes can be challenging to compare in the presence of stimulus symmetries. 

\begin{figure}[t!]
    \centering
    \includegraphics[width=4in]{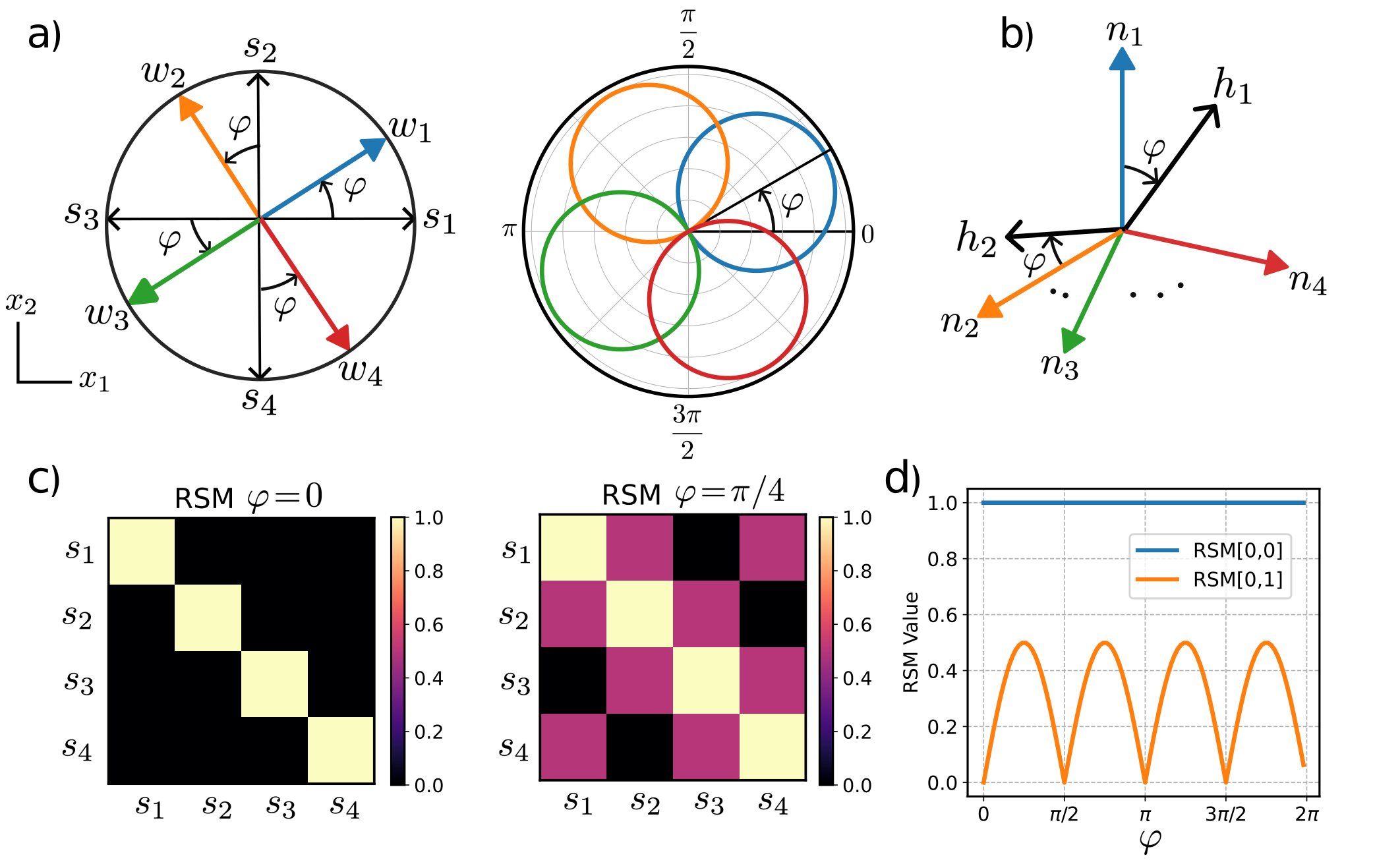}
    \vspace{-1em}
    \caption{\textbf{Gauge-dependent representations in a toy model.} a) Left: RF center vectors ($w_i$) of four neurons tiling a one-dimensional ring. To specify a tiling, one must choose a global orientation $\varphi$. Right: corresponding angular tuning curves. b) Representations $h_{1}$ and $h_{2}$ of two trial stimuli $s_1$ and $s_2$ in the four-dimensional representation space. The angle between $h_1$ and $h_2$ depends on $\varphi$. c) RSMs for two values of the gauge variable $\varphi$. d) RSM diagonal and off-diagonal entries as a function of $\varphi$. The variation of the RSM with $\varphi$ is the key phenomenon which this work explores.}
    \label{fig1}
    \vspace{-1em}
\end{figure}

\section{Gauge-dependent RSMs in a toy model}\label{sec:toy}

We begin with a toy example that illustrates the key phenomenon, motivated by classic models for orientation encoding in neuroscience \citep{kriegeskorte2021geometry}. Consider a setting where data lies on a one-dimensional ring ($x \in \mathbb{S}^1$), encoded by $n$ neurons with localized receptive fields that tile this space uniformly. Considering this tiled representation, we see there is an ambiguity: given a particular spacing between receptive fields, there exist many possible tilings of the stimulus manifold which are functionally equivalent given the rotational symmetry. Therefore, to completely specify the representation, one must fix a gauge; one must anchor the lattice of receptive fields (Fig.~\ref{fig1}a).\footnote{Our use of the term ``gauge'' is inspired by physics, though the phenomenon we describe is not identical, as here the gauge is observable but not functionally relevant \citep{jackson2001gauge}. }

Concretely, the activation of the $i$-th neuron in response to an input $x$ is 
\begin{equation}
    h_{i}(x) = \operatorname{ReLU}(w_{i}^{\top} x), \quad \textrm{where} \quad w_{i} = (\cos\theta_i,\sin\theta_i) \quad \textrm{for} \quad \theta_{i} = 2(i-1)\pi/n + \varphi. 
\end{equation}
That is, each neuron has a unit-length weight vector $w_{i}$, and a preferred orientation $\theta_{i}$. Here, $\varphi$ is an arbitrary offset angle which defines the global configuration; it is the gauge variable. Since each neuron has a receptive field (RF) that covers half the space ($\pi$ radians on the ring), $n=4$ is the minimum number of neurons that can cover this space faithfully (Fig. \ref{fig1}a; see App.~\ref{AppendixA}). To be precise, for any $\varphi$, one can perfectly recover $x \in \mathbb{S}^{1}$ from $h(x)$ using a linear readout whose weights are rotated to compensate for the change in the representation as $\varphi$ varies (App.~\ref{readout}).

In this configuration, one can track the representations of a set of trial stimuli, $s_1=(1,0)^{\top}, s_2=(0,1)^{\top}$, $s_3=(-1,0)^{\top}$ and $s_4=(0,-1)^{\top}$.\footnote{We use these trial stimuli to show a compact form of the RSM; for arbitrary stimuli in $\mathbb{S}^{1}$ see App.~\ref{arbitraryanglen=4} and Fig.~\ref{fig_stim_supp}.} Collecting the representations into a matrix $H = [h(s_1),h(s_2),h(s_3),h(s_4)]$ and defining the associated $RSM = H^{\top} H$, we have
\begin{align}
    H = 
    \begin{pmatrix}
        \cos(\varphi) & \sin(\varphi) & 0 & 0\\
        0 & \cos(\varphi) & \sin(\varphi) & 0 \\
        0 & 0 & \cos(\varphi) & \sin(\varphi) \\
        \sin(\varphi) & 0 & 0 & \cos(\varphi)
    \end{pmatrix} \quad\textrm{ and }\quad
    \label{RSMtoy}
  RSM = \begin{pmatrix}
1 & \rho(\varphi) & 0 & \rho(\varphi) \\
\rho(\varphi) & 1 & \rho(\varphi) & 0 \\
0 & \rho(\varphi) & 1 & \rho(\varphi) \\
\rho(\varphi) & 0 & \rho(\varphi) & 1
\end{pmatrix}
\end{align}
for $\varphi \in [0,\pi/2)$, where $\rho(\varphi) = \sin{\varphi}\cos{\varphi}$. Increasing $\varphi$ beyond $\pi/2$ is equivalent to circularly permuting the neuron labels and taking the residue of $\varphi$ modulo $\pi/2$. Thus, the RSM is clearly gauge-dependent (Fig.~\ref{fig1}c-d). This change is qualitatively substantial, as by changing $\varphi$ one can achieve either a purely diagonal RSM or one where the off-diagonal elements are as large as one-half.

Therefore, in a simple case where changing the gauge variable $\varphi$ corresponds to an overall rotation of the tuning curves, the RSM depends on the choice of gauge. This results from the fact that the global rotation of the tuning curves does not translate to an overall rotation in the representation space; the angle between the representations of two stimuli can change (Fig. \ref{fig1}b).

\section{Formalizing RSM gauge-dependence for general data symmetries}\label{sec:formalism}

The toy model in \S\ref{sec:toy} illustrates that two notions of representational equivalence need not coincide. First, representations related by a gauge transformation are functionally equivalent from the perspective of any downstream computation that respects the stimulus symmetry. Second, representations related by an orthogonal transformation have identical RSMs. The toy model shows that gauge transformations need not induce orthogonal transformations of the representation, meaning that \textit{equivalence up to a gauge does not imply equivalence of RSMs}. This distinction generalizes beyond the toy model, as we show formally below. We note that this formalism is not essential for understanding our later results. 

\subsection{Symmetries of the stimulus distribution}

Consider a $d$-dimensional space of stimuli $X$, distributed according to a probability measure $\mu$. Suppose that some compact group $G$ acts on $X$, via $\cdot : G \times X \to X$. In general, $G$ defines a symmetry of the stimulus if the distribution $\mu$ is invariant under this action, \textit{i.e.}, if for every measurable set $A \subseteq X$ and any $g \in G$ we have
\begin{align}
    \mu(g A) = \mu(A). 
\end{align}
In many applications, this symmetry arises from a lower-dimensional latent variable. For example, rotations of an image can be parameterized by a latent orientation $z \in \mathbb{S}^{1}$, even though the observed stimuli themselves live in a high-dimensional pixel space (Figure~\ref{fig_man_schem}). The arguments below apply to stimuli with such latent symmetries; all we require is an induced action of the symmetry group on the stimulus distribution. 

\subsection{Gauge transformations and functional equivalence}\label{sec:decoding}

We now encode the data into an $n$-dimensional representation. Most simply, the encoding could be defined by a deterministic encoding function $h : X \to H \subseteq \mathbb{R}^{n}$. However, representations in biological systems are noisy: neural responses to the same stimulus vary considerably across trials, even under controlled experimental conditions \citep{pouget2000information,goris2014partitioning}. With this in mind, we consider instead a general stochastic encoding channel $\rho$, which given a stimulus $x \in X$ assigns probability $\rho(A\mid x)$ to measurable subsets $A \subseteq H$ of the representation space. This covers deterministic encodings as the special case $\rho(\cdot \mid x) = \delta_{h(x)}(\cdot)$ in which $\rho(du\mid x)$ is a point mass at $h(x)$. 

We can define the action of the symmetry group $G$ on the encoding by its action on the inputs: for any $g \in G$, 
\begin{align}\label{eqn:symm_noise}
    (g\cdot\rho)(A\mid x) = \rho(A \mid g^{-1} \cdot x). 
\end{align}
We refer to this as a \emph{gauge transformation}. In the simple case of a deterministic encoding, a gauge transformation is just $(g\cdot h)(x) = h(g^{-1} \cdot x)$. Importantly, gauge transformations are very far from arbitrary transformations of the representation: they relabel which stimulus is associated to which distribution of responses according to the symmetry. 

We now show that gauge transformations preserve downstream computational functions that respect the symmetry. Consider a general task defined by a mapping $t : X \to Y$ that maps each stimulus to a target in some space $Y$. Suppose that $G$ also acts on $Y$, and that the task is equivariant under the symmetry:
\begin{align}
    g \cdot t(x) = t(g \cdot x).
\end{align}
We measure performance on the task via a $G$-invariant loss function $\ell : Y \times Y \to \mathbb{R}$, \textit{i.e.}, one that satisfies
\begin{align}
    \ell(g \cdot y, g \cdot y') = \ell(y,y')
\end{align}
for all $y,y'\in Y$ and any $g \in G$. Together, these conditions give a general formal definition of downstream functions that respect the stimulus symmetry. 

Now, suppose that we try to use a given representation $\rho$ to solve this task, via a readout function $f : H \to Y$. The action of $G$ on $Y$ allows us to define the transformed readout $(g \cdot f)(u) = g \cdot f(u)$ for any $u \in H$. On average over the randomness in the encoding, this gives us an expected task loss
\begin{align}
    \mathcal{E}[\rho,f] = \mathbb{E}_{x\sim\mu} \mathbb{E}_{u \sim \rho(\cdot\mid x)}[\ell(t(x),f(u))]. 
\end{align}
Under the stated assumptions, this task loss is gauge-invariant: for any $g \in G$, any decoder $f$, and any encoding $\rho$, we have
\begin{align}
    \mathcal{E}[g \cdot \rho, g \cdot f] = \mathcal{E}[\rho,f]. 
\end{align}
The proof of this claim essentially follows by a change of variables, so we give it in the main text below:
\begin{align}
    \mathcal{E}[g \cdot \rho, g \cdot f] 
    &= \int_{X} \int_{H} \ell(t(x),g\cdot f(u)) \rho(du\mid g^{-1}\cdot x) \mu(dx) && (\textrm{def. of } g \cdot \rho)
    \\
    &= \int_{X} \int_{H} \ell(t(g\cdot x'),g\cdot f(u)) \rho(du\mid x') \mu(dx') && (x' = g^{-1}\cdot x, \textrm{ $G$-inv. of } \mu)
    \\
    &= \int_{X} \int_{H} \ell(t(x'),f(u)) \rho(du\mid x') \mu(dx') && (\textrm{$G$-eqiv. of $t$ \& $G$-inv. of $\ell$})
    \\
    &= \mathcal{E}[\rho,f] && (\textrm{def. of } \mathcal{E}[\rho,f]). 
\end{align}
This is a strong functional equivalence result: for any task that respects the stimulus symmetry, the performance on that task as measured by a $G$-invariant loss is exactly invariant under gauge transformations of the representation. An even stronger gauge invariance result holds for $G$-invariant tasks, \textit{i.e.}, those with $t(g \cdot x) = t(x)$. Then, no transformation of the readout is necessary, as if $t$ is invariant then the full joint distribution of $(u,t(x))$ is gauge-invariant. The proof of this claim is essentially the same as the argument applied to equivariant tasks above; for completeness we give in Appendix~\ref{app:formal}. 

This functional equivalence applies only to tasks that respect the symmetry. If a downstream computational task itself breaks the symmetry, then the gauge is no longer functionally-irrelevant with respect to that task. For example, in the case of recognizing a rotated image, it may be the case that the four cardinal orientations have distinct significance relative to the other orientations. Then, the gauge is fixed by the task structure. 

As we did in the toy model of \S\ref{sec:toy}, we will often focus on autoencoding as a model problem. In the setup above, autoencoding corresponds to setting $Y=X$ with the same action of $G$ as on $X$, $t=x$, and $\ell(x,x') = d_{X}(x,x')$, where $d_{X}$ is a general $G$-invariant metric on $X$. Thus, autoencoding performance is gauge-invariant so long as we measure reconstruction accuracy with an invariant distance metric. 

\subsection{Gauge dependence of the RSM}\label{sec:rsmgauge}

Though the decoding accuracy is gauge-invariant, the encodings themselves are clearly not in general gauge-invariant. This gauge-dependence will in general carry over to the RSM, as it is invariant only if the gauge transformation preserves pairwise inner products in representation space. Formally, for a deterministic encoding $h(x)$, the RSM is defined for $x,x' \in X$ by the Euclidean inner product of $h(x)$ and $h(x')$: 
\begin{align}
    RSM_{h}(x,x') = h(x)^{\top} h(x') = \sum_{i=1}^{n} h_{i}(x) h_{i}(x') . 
\end{align}
For a stochastic encoding, one could either compute the RSM from the conditional mean response
\begin{align}
    h_{\rho}(x) = \mathbb{E}_{u \sim \rho(\cdot \mid x)}[u], 
\end{align}
which is the standard neuroscience definition of a tuning curve or receptive field in the presence of noise or nuisance variables \citep{goris2014partitioning,onoo2026readout}, or from individual noisy measurements. Though recent works have considered how to define noise-aware similarity measures \citep{duong2023representational}, we will focus on the mean responses, which we will write as $h$ rather than $h_{\rho}$ for brevity.

For the RSM to be gauge-invariant, we must have 
\begin{align}
    RSM_{g \cdot h}(x,x') = RSM_{h}(x,x')
\end{align}
for all $x,x' \in X$ and any $g \in G$. As we show formally in Appendix~\ref{app:formal}, this holds if and only if there exists an orthogonal matrix $O \in \mathcal{O}(n)$, depending on the group element $g$ but not on $x$, such that 
\begin{align}
    (g\cdot h)(x) = O(g) h(x).
\end{align}
This result follows from the classical fact that two sets of vectors having the same Gram matrix must be related by an orthogonal transformation \citep{horn2012matrix}. 

\begin{figure}[t]
    \centering
    \includegraphics[width=5.8in]{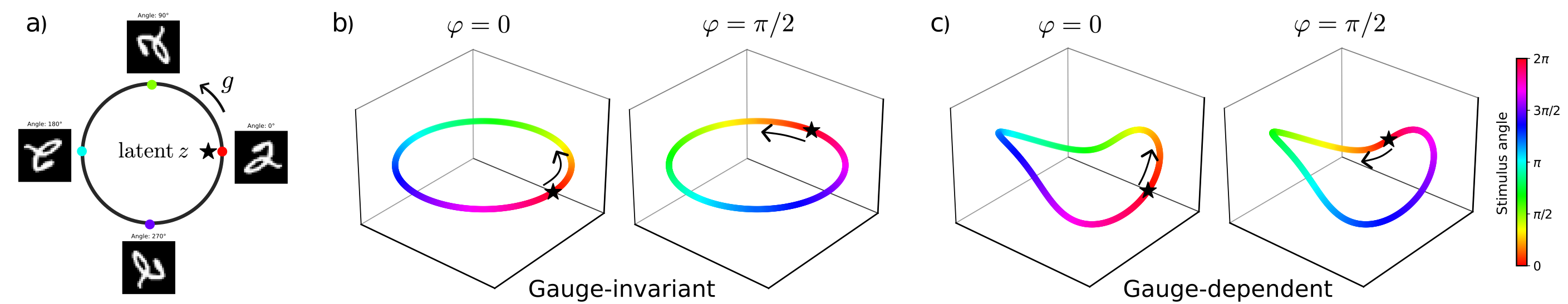}
    \caption{\textbf{Gauge dependence and the geometry of representational manifolds.} The schematics show a) the latent space $Z = \mathbb{S}^1$, and two examples of b) linear and c) nonlinear embeddings of it in $\mathbb{R}^3$. Points are color-coded by the stimulus angle, and angle zero is marked by star $(\star)$. The left and right columns have a gauge difference of $\Delta \varphi = \pi/2$. The two manifolds in (b) can be transformed into each other using an orthogonal transformation, but this is not possible in (c). Consequently, per Section~\ref{sec:rsmgauge}, the representations in (c) have gauge-dependent RSMs.}
    \label{fig_man_schem}
    \vspace{-1em}
\end{figure}

In group-theoretic terms, this would mean that the encoding is an orthogonal linear representation of $G$ \citep{hall2003lie}. This is satisfied by construction for equivariant architectures \citep{bronstein2021geometric,cohen2015transformation,kondor2018convolution}. However, as we saw in \S\ref{sec:toy}---and as we will see in the subsequent sections of the paper---this is not in general the case, even for simple and biologically-reasonable nonlinear encodings. We illustrate the geometry underlying the gauge-variance of the RSM in Figure \ref{fig_man_schem}: the representational manifolds before and after a gauge transformation need not be related by a rotation. 

The analysis above considers invariance at the level of the full RSM kernel, \textit{i.e.}, for \textit{any} pair of stimuli $x,x' \in X$. In practice, one can test only a finite set of trial stimuli, as we did for the toy model in \eqref{RSMtoy} (though, see Fig.~\ref{fig_stim_supp}). Formally, showing gauge dependence of the RSM for a finite set of trial stimuli is sufficient to show that the representation is not orthogonally equivariant. However, equality of the finite RSMs is not in general sufficient to show orthogonal equivariance on the full stimulus space (see Appendix~\ref{app:formal}). In principle, the choice of trial stimuli could either mask or exaggerate the gauge-dependence of the full RSM.

\subsection{Fixed noise channels preserve functional equivalence}

Our results above show that noise \emph{per se} cannot break the functional symmetry. However, this abstract result is somewhat unrealistically permissive, as it assumes that we are free to define the group action \eqref{eqn:symm_noise} on the stochastic encoding channel. A more reasonable model for noisy population encoding is to specify an underlying deterministic representation together with a fixed noise model \citep{pouget2000information,goris2014partitioning}. Then, the question is whether changing the gauge of the deterministic representation while holding the noise model fixed leads to the action \eqref{eqn:symm_noise}. 

This property is satisfied by any noise model that depends on $x$ only through the deterministic response $h(x)$, which includes many standard models for neural noise \citep{pouget2000information}. Concretely, suppose that there is a fixed channel $\nu$, independent of $h$, such that $\rho_{h}(A\mid x) = \nu(A \mid h(x))$. Then, $\rho_{g \cdot h}(A \mid x) = \nu(A \mid h(g^{-1} \cdot x)) = \rho_{h}(A\mid g^{-1} \cdot x)$, which is the desired transformation law \eqref{eqn:symm_noise}. One such noise model is fixed additive noise, \textit{i.e.}, if the observations are given by $h(x) + \epsilon$ for $\epsilon$ drawn from any fixed zero-mean distribution that is independent of $x$ and of the gauge. A more interesting example is the classic model of Poisson noise on top of fixed tuning curves, as then the noise distribution is signal-dependent. Concretely, starting from the toy model of \S\ref{sec:toy}, suppose that the $i$-th neuron is characterized by a noisy conditional response $r_{i} \mid \theta,\varphi \sim \mathrm{Poisson}(T h_{i}(\theta,\varphi))$ independent across neurons, where $T$ is an observation window and $h_{i}(\theta,\varphi) = h(\theta-\theta_{i}(\varphi))$ is the tuning curve defined by the same model as in \S\ref{sec:toy}, with gauge-dependent center location $\theta_{i}(\varphi)$. Then, the population response distribution is given by a conditional measure $\rho_{\varphi}(r\mid \theta)$, depending on the gauge angle $\varphi$ and stimulus angle $\theta$, that satisfies $\rho_{\varphi + \delta}(r \mid \theta) = \rho_{\varphi}(r\mid\theta-\delta)$ for any angular shift $\delta$. This property is an instance of the action \eqref{eqn:symm_noise} for stimuli on the circle. 

Therefore, noise does not generically render gauge transformations functionally-relevant. It does so only when the noise model itself introduces symmetry-breaking structure, for instance, if its statistics depend directly on absolute stimulus coordinates or on the representation's gauge in a manner not mediated by the underlying deterministic response. 

\section{Towards real data: factors influencing gauge dependence in the toy model}\label{sec:factors}

\begin{figure}[t]
    \centering
    \includegraphics[width=5.7in]
    {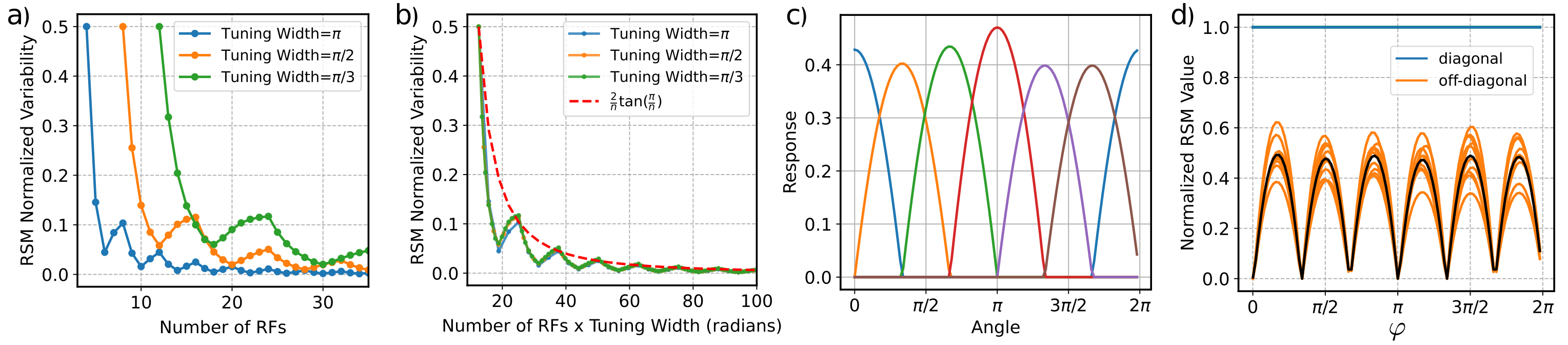}
    \caption{\textbf{Factors influencing RSM gauge dependence.} a,b) RSM variability ($\Delta$) as a function of number of RFs and tuning width. In (b), the dashed red line shows the fixed-width prediction $\Delta = \frac{2}{n} \tan\left(\frac{\pi}{n}\right)$.
    c) RFs with amplitude noise (additive Gaussian noise with standard deviation $0.1$). d) Corresponding $\varphi$-dependence of RSM over 100 realizations of noisy RFs (black line: average).
    }
    \label{fig2}
    \vspace{-1em}
\end{figure}

In \S\ref{sec:toy}, we showed gauge-dependence of the RSM in a highly idealized toy model. With the objective of demonstrating that this phenomenon is present in networks where the RFs are learned rather than hand-designed in mind, we build intuition by considering how introducing various departures from this toy model affect the gauge-dependence of the RSM. Specifically, we will study the effects of the number, the width, and the amplitude of the RFs (see App.~\ref{sec:appendix:toysetup} for detailed methods). 

We quantify the $\varphi$-dependence of the RSM by the range of its off-diagonal elements: $\Delta := \max_{\varphi}(RSM_{off})-\min_{\varphi}(RSM_{off})$. Here, $RSM_{off}$ is an off-diagonal element of the normalized RSM, or cosine similarity matrix; normalization does not remove the $\varphi$-dependence. $\Delta$ depends on the choice of trial stimuli. Unless otherwise stated, we calculate $\Delta$ for stimuli that are separated by half of the tuning width, which is equivalent to the maximum $\Delta$ among all possible trial stimuli. 

For a given tuning width, increasing the number of RFs suppresses variability in the RSM (Fig.~\ref{fig2}a-b). However, the product of tuning width and the number of RFs matters in this relationship: for small tuning widths, the variability $\Delta$ is non-negligible even if the number of RFs is large (Fig.~\ref{fig2}a-b and \ref{fig_phicurves_supp}). In Appendix \ref{AppendixA}, we show that for $n$ a multiple of 4 and a tuning width of $\pi$, the RSM has a simple closed-form expression which generalizes that found for $n=4$; all that changes is the constant value of the diagonal elements and the details of the function $\rho(\varphi)$. For $n$ a multiple of four, this leads to $\Delta = \frac{2}{n} \tan\left(\frac{\pi}{n}\right)$, meaning that the variability in the RSM vanishes as $\mathcal{O}(n^{-2})$ for fixed tuning width. This is plotted as a dashed red line in Fig.~\ref{fig2}b. However, this does not hold if tuning width scales with the number of receptive fields. Changing the width of the RFs is approximately equivalent to re-scaling the input (App.~\ref{sec:appendix:toysetup}), which explains the near-collapse of curves in Fig.~\ref{fig2}b. This implies that even at a high number of neurons, decreasing the tuning width will accentuate the gauge-dependence.

Finally, we studied the effect of noisy RFs on the RSM variability. Fig.~\ref{fig2}c,d show an example of RFs with amplitude noise and the corresponding gauge dependence curves for multiple runs. Despite the noise, the expected modulation by $\varphi$ is noticeable, especially in the average (black curve; additional results in Fig.~\ref{fig_amp_supp}). 

\section{Learning-induced manifold-tiling and drift in representations of the circle} \label{sec:learning}

\subsection{SGD need not select a unique RSM}

So far, we showed that a manifold-tiling solution can create a varying RSM as a function of the functionally-irrelevant gauge variable, provided that the tiling is not too dense. We now study this in a learning scenario, focusing on two-layer autoencoders with ReLU neurons trained using SGD with weight decay on isotropic $d$-dimensional Gaussian data. We give more details and additional experimental results in Appendix \ref{app:experiments}. 

We first replicate our toy model results, showing that SGD training replicates the uniform-tiling solution for the $n=4$ neurons in the hidden layer. Over the course of continual training, the tuning curves and the RSM drift over time, without selecting a particular solution (Fig.~\ref{fig_sgd_242}). This ambivalence also holds across different realizations of the learning started from different initializations of the weights (Fig.~\ref{fig_supp_init_2_4_2}).
\begin{figure}[t!]
    \centering
\includegraphics[width=5.5in]{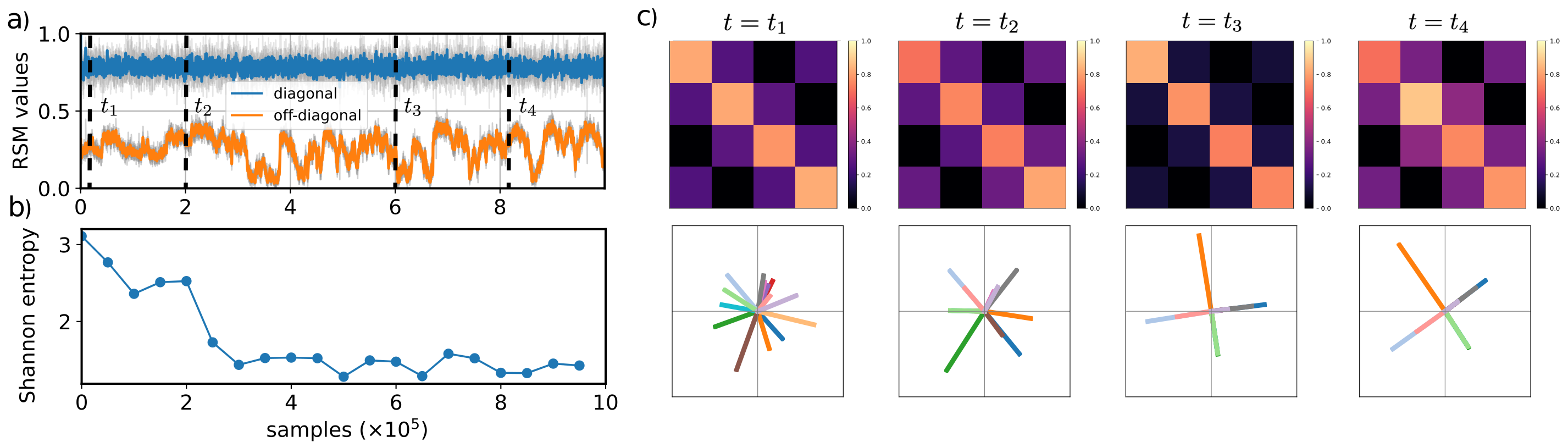}
    \caption{\textbf{Collapse of neurons and RSM drift under continual SGD training.}
    a) RSM values over time.
    b) Entropy of the distribution of cosine of pairwise angles between all neurons' weights. c) RSMs (top), and the corresponding weight vectors (bottom) at different snapshots during training. Alignment of neurons' weights $(n=15)$ into 4 orthogonal directions is evident through learning.}
    \label{fig4}
\end{figure}

\subsection{SGD inductive bias or energetic regularization can generate sparsely-tiling solutions}

\begin{figure}[h!]
    \centering
    \begin{minipage}[t]{2.25in}
        \vspace{0pt}
        \centering
        \includegraphics[width=\linewidth]{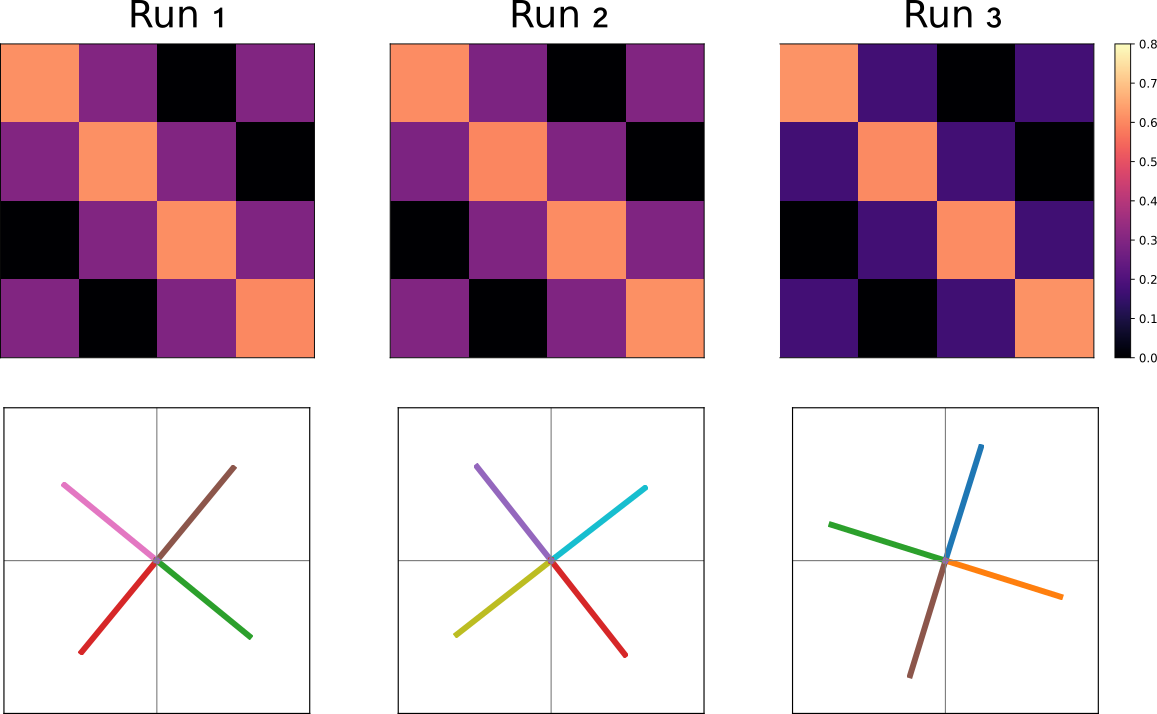} 
    \end{minipage}\hfill\begin{minipage}[t]{4in}
    \vspace{0pt}
    \centering
    \vspace{0.75em}
    \caption{\textbf{Gauge dependence of RSM under an $L_1$ penalty on activations.} RSM plots (top) and neurons' weight vectors (bottom) for three different runs ($n=15$ neurons). Similar to Figure \ref{fig_supp_init_2_15_2}, the batch size is large but the $L_1$ penalty on the hidden-layer activations leads to all but 4 neurons remaining active. Each simulation is run with $5 \times 10^4$ samples seen (batch size of 100).}
    \label{fig_supp_init_2_15_2_l1}
    \end{minipage}
\end{figure}

We found in \S\ref{sec:factors} that for a fixed tuning width, densely tiling the manifold with many neurons suppresses variability in the RSM. Here, we demonstrate two ways that learning can lead to solutions with highly-variable RSMs, even when the number of hidden neurons is large. First, SGD can align the receptive fields of multiple neurons, leading to a solution with \textit{effectively} only four neurons (Fig.~\ref{fig4}). After this collapse, the gauge continues to drift, and the RSM varies accordingly (see panels $t=t_3$ and $t=t_4$ in Fig.~\ref{fig4}c). This collapse can occur when SGD noise is large; see Fig. \ref{fig_supp_init_2_15_2} for an example of a low-noise regime. Second, regularizing the activations with an $L_1$ penalty---to mimic energetic costs---can lead to all but four neurons becoming silent (see Fig. \ref{fig_supp_init_2_15_2_l1}). This effect also allows RSM variability to exist despite a high number of neurons. Due to the fact that data are sampled uniformly from the latent space, the $L_1$ penalty does not fix the gauge (App.~\ref{l1_gauge}).

\section{Structured RSM variability for higher-dimensional spherical stimuli} \label{sec:high-dim}

We have thus far considered a two-dimensional stimulus with a one-dimensional symmetry group (rotations in the plane). However, in keeping with the general formalism of \S\ref{sec:formalism}, the same gauge redundancy is present in representations of higher-dimensional symmetric manifolds. Analytical study of tiled representations of higher-dimensional manifolds is challenging because the maximally uniform arrangement of receptive field centers is generally unknown, even for the sphere $\mathbb{S}^{2}$ \citep{delbono2024most}. However, if we tile the hypersphere $\mathbb{S}^{d-1}$ with a reflection-symmetric arrangement of neurons---grouped in pairs with oppositely-oriented receptive fields---we can predict the structure of variability in the RSM induced by the $d(d-1)/2$-dimensional continuous $\mathrm{SO}(d)$ gauge symmetry.

\begin{figure}[t!]
    \centering
\includegraphics[width=0.7\linewidth]{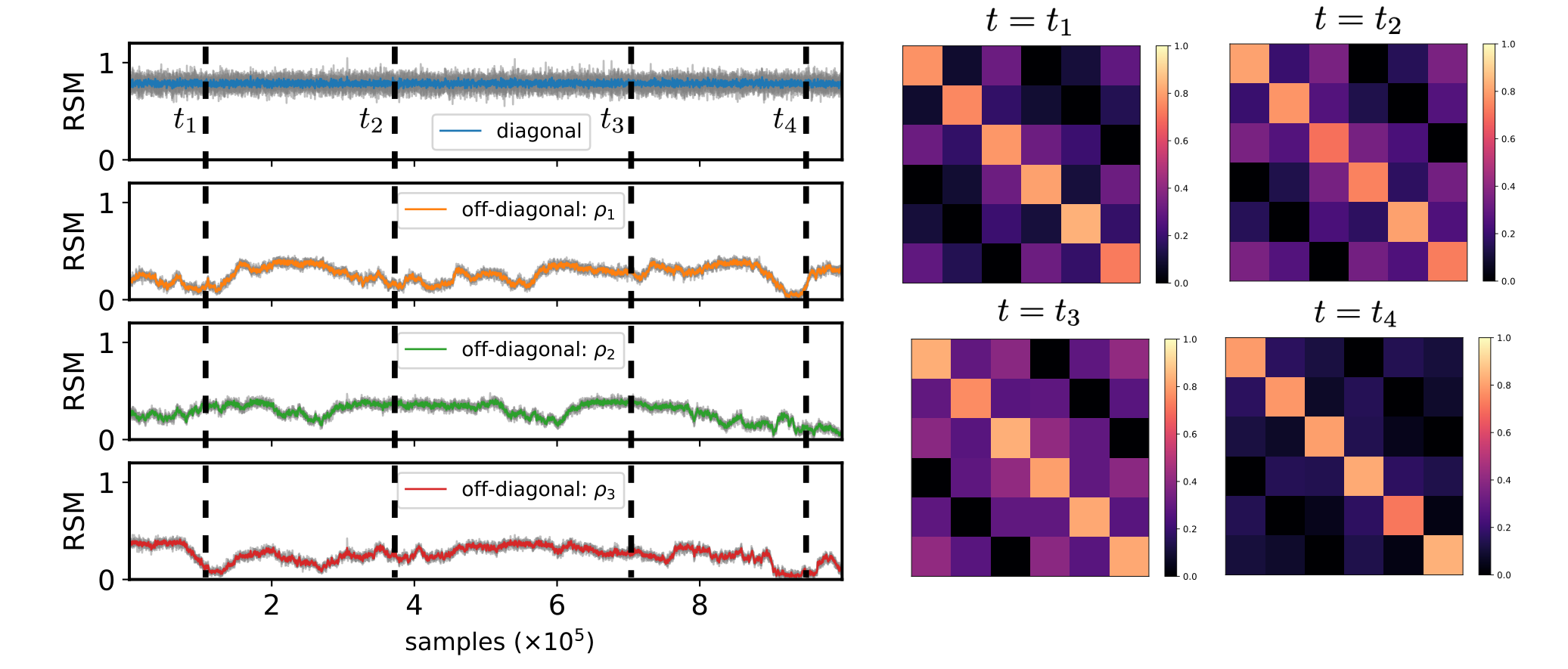} \caption{\textbf{RSM variability as a result of continual training for three-dimensional input.} (Left) Components of the empirical RSM as a function of time. The non-zero off-diagonal elements are grouped into three distinct sets ($\rho_1$, $\rho_2$, and $\rho_3$) based on the predictions derived in Appendix \ref{app:sphere} (this implies entries $[0,1]$,$[0,4]$,$[1,3]$, $[3,4]$, and their transpose vary as $\rho_1$, and similarly for other groups). Gray curves correspond to the group members and the colored curves denote the group averages. (Right) RSM matrices at different snapshots demonstrate variability through time. There are $n=6$ ReLU neurons in the network. }
\label{fig_supp_3_6_3}
\end{figure}

Concretely, we consider an encoding $h(x) = \ReLU(W^{\top} x)$, where the weight matrix $W = (U^{\top},-U^{\top})$ for some $U \in \mathbb{R}^{n/2 \times d}$. Now, the test stimuli are the $d$ cardinal basis vectors and their negations. To perfectly recover the test stimuli using tied decoder weights, the matrix $U$ must be semi-orthogonal, \textit{i.e.}, $U^{\top} U = I_{d}$. For any such semi-orthogonal $U$, the RSM evaluated on the test stimuli is given by 
\begin{align}
    RSM = \begin{pmatrix} I_{d} + R & R \\ R & I_{d} + R \end{pmatrix},
\end{align}
where $R$ is a $d \times d$ symmetric matrix with zeros along the diagonal, which thus has $d(d-1)/2$ independent elements (see derivation in App.~\ref{app:sphere}). In three dimensions, there are therefore three sets of distinct non-zero off-diagonal elements of the RSM, each of which varies over time during continual training (Fig. \ref{fig_supp_3_6_3}). Not only is the RSM highly structured in the sense that its elements cluster into groups, but the variability of these groups is correlated (see also Fig. \ref{fig_10_20_10} for $d=10$, where there are $d(d-1)/2 = 45$ distinct groups). In the simulations, we use isotropic Gaussian data, meaning that they include a radial nuisance variable. 

\section{RSM variability with latent symmetries in image data}\label{sec:real}

\begin{figure}[t!]
    \centering
    \includegraphics[width=5.5in]{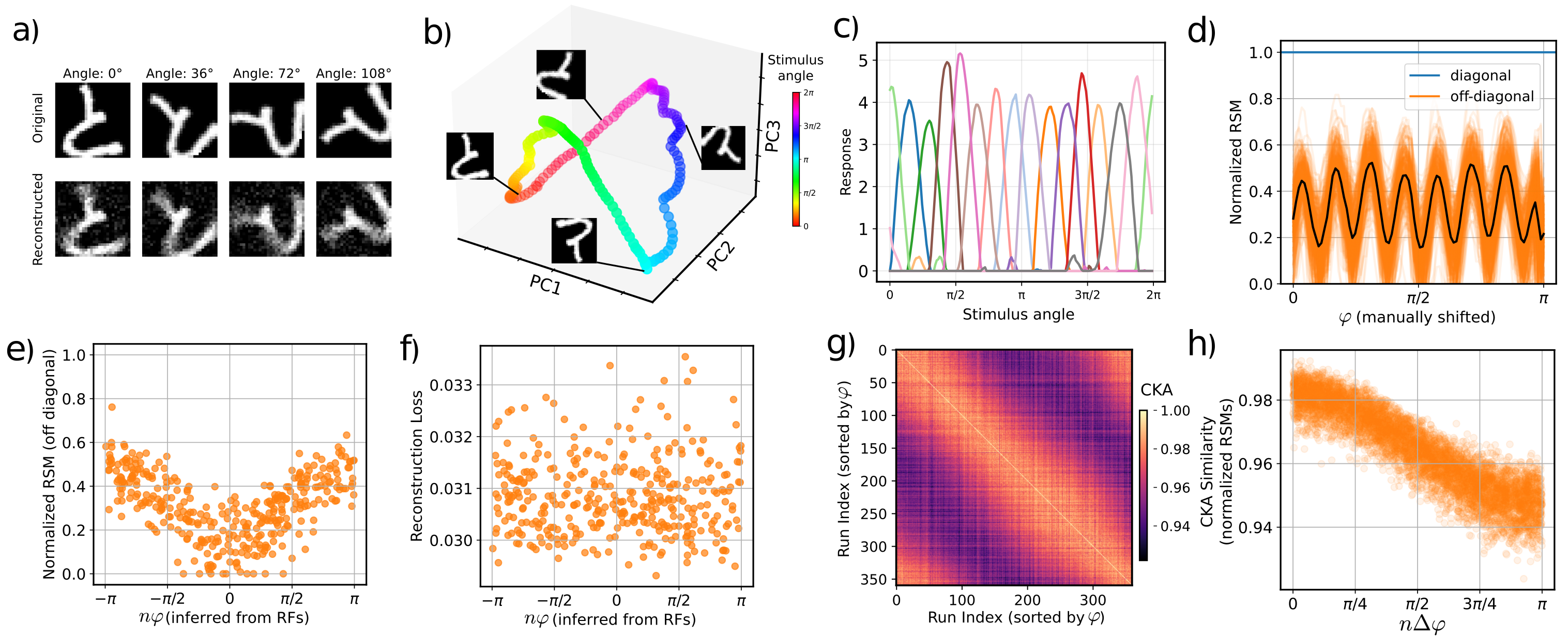}
    \caption{\textbf{RSM variability in autoencoding a rotated image manifold.} Autoencoders were trained on rotated versions of a digit from Kuzushiji-MNIST dataset compiled by \citet{clanuwat2018deep}. a) Four examples of the original and reconstructed images in an instance of the trained model. b) 3D PCA of the hidden layer activations in a trained model in response to all stimuli. Each point is color-coded by the rotation angle of the stimulus. c) RFs in a model with $n=15$ active neurons. d) $\varphi-$dependence curves of RSM. Each orange curve corresponds to a run and is a result of manually shifting the RFs. Curves are aligned and the black curve shows the average. 
    e) Scatter plot of normalized off-diagonal RSM vs. the RF-inferred $\varphi$ across different trained networks. Each point shows a different run of the model that had $n=15$ active RFs (see Fig.~\ref{fig_realdata_supp} for different $n$'s). The off-diagonal RSMs are calculated for two stimuli that are $\alpha = 25^{\cdot}$ apart.
    f) Same as (e) but the y-axis is the reconstruction loss for each run. g) CKA similarity matrix calculated between normalized RSMs of different runs. The runs are sorted based on the inferred $\varphi$. h) Scatter of CKA similarity as a function of gauge difference (each point represents a pair of runs, and for visualization purposes only $1/10$ of the data are shown).
    }
    \label{fig_real}
\end{figure}

So far, we have demonstrated RSM variability in networks trained on Gaussian data. Here, we consider a more realistic case where the data are generated by rotations of a static image from the Kuzushiji-MNIST dataset \citep{clanuwat2018deep}. The latent space is $\mathbb{S}^1$, corresponding to stimuli with different rotation angles, but the stimulus space is more complicated than previous cases.  We train a two-layer autoencoder with ReLU nonlinearity on 180 rotated versions of the image (Fig.~\ref{fig_real}a; see App.~\ref{app:real_methods} for model details). Localized and noisy RFs are formed in the hidden layer, and they approximately tile the $\mathbb{S}^1$ latent space (Fig.~\ref{fig_real}b,c).

To study the gauge dependence of the RSM, we train 1000 model realizations and perform the following two analyses. First, similar to the previous $\varphi$-dependence curves in the toy models (e.g. in Fig.~\ref{fig2}d), for each trained network we manually change the gauge by circularly shifting the RFs and compute corresponding RSMs. As shown in Fig.~\ref{fig_real}d, the $\varphi$-dependence curves oscillate at intervals of $2\pi/n$, consistent with the toy model results. Second, since the models are trained independently, we expect them to have randomly distributed gauge $\varphi$. For each run, we estimate $\varphi$ by Fourier-transforming the envelope of the RF curves, and extracting the phase associated with the dominant frequency (see App.~\ref{app:real_methods} and Fig.~\ref{fig_realdata_method}). We then plot a scatter of the normalized off-diagonal element of the RSM as a function of this inferred $\varphi$ in Fig.~\ref{fig_real}e. The U-shape visible in this plot confirms the expected gauge dependence of the RSM across the models (Pearson's $r = 0.76, p<0.001$, between the RSM and absolute phase). Importantly, the reconstruction loss shows no clear $\varphi$-dependence (Fig.~\ref{fig_real}f, Pearson's $r = 0.03, p=0.56$; see additional results in Fig.~\ref{fig_realdata_supp}). 
Finally, we show that the CKA similarity drops significantly as a function of the difference in the gauge variable (Fig.~\ref{fig_real}g-h, Pearson's $r=-0.91, p<0.001$). These results show that, in a more realistic setup with noisy RFs, the gauge dependence of the RSM can be distinguished from other sources of RSM variability.

\begin{figure}[t!]
 \centering
\includegraphics[width=5.0in]{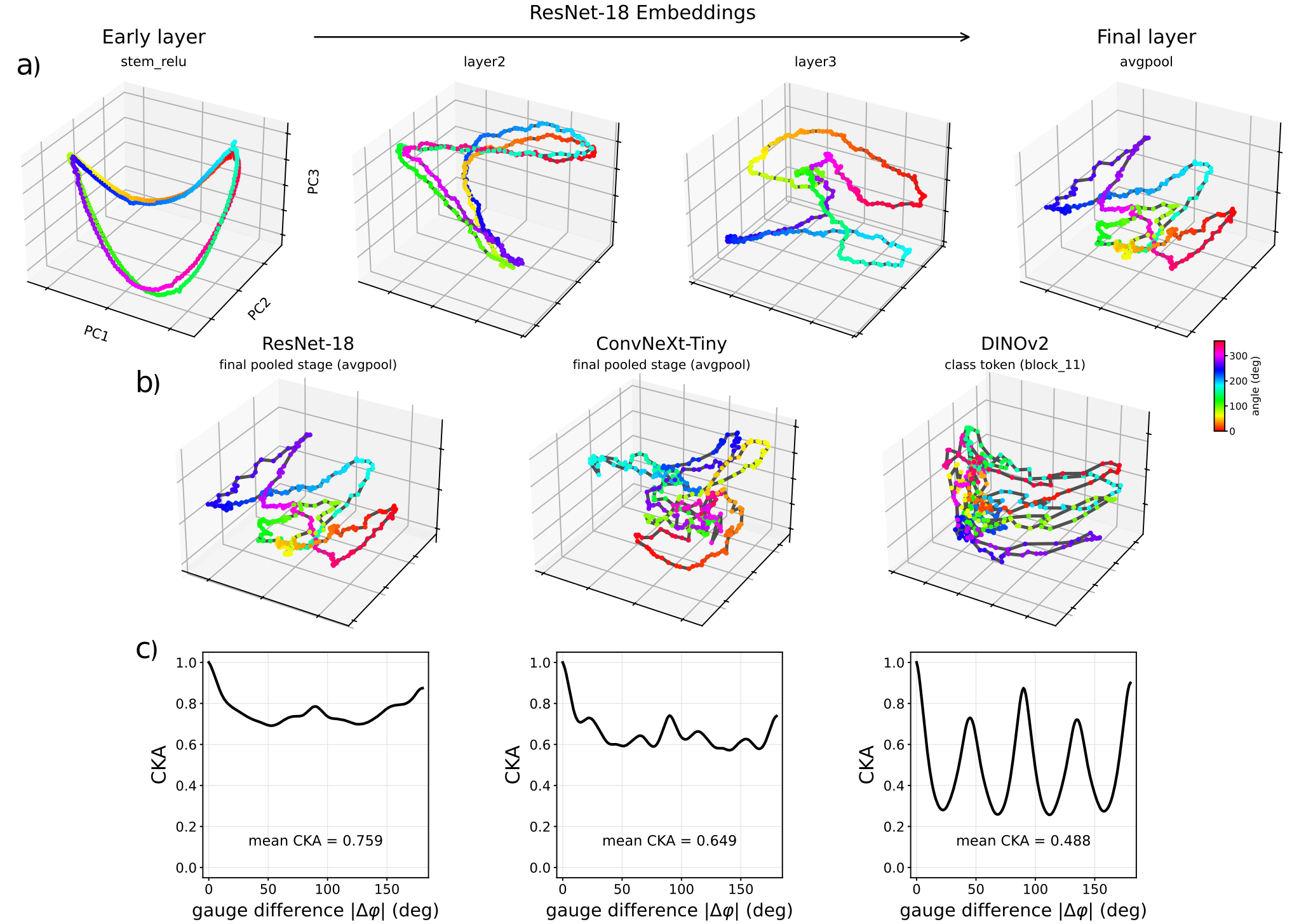}
    \caption{Representations of rotated images 
    in pretrained vision models. a) 3D PCA embeddings across layers of ResNet-18. Color denotes rotation angle. b) Embeddings in the final layers of three exemplary pretrained models: (left) ResNet-18 \citep{he2016deep}, (middle) ConvNeXt-Tiny \citep{liu2022convnet}, and (right) DINOv2-S/14-reg \citep{oquab2023dinov2}. The first two models are convolution-based, and the last model is transformer-based. c) CKA similarity as a function of gauge difference for the three models in (b). Note that here, different gauge values were created manually by offsetting the RSM according to the starting angle. This corresponds to circularly shifting the rows and columns of RSM simultaneously.}
\label{fig_realdata_largemodel}
\end{figure}

As a final illustration of how latent rotational symmetries in image data can affect RSM-based analyses, we consider several standard pretrained general-purpose image models \citep{he2016deep,liu2022convnet,oquab2023dinov2}. Previous works have shown that image models not constrained to obey rotation-equivariance generally do not learn this symmetry from data \citep{lenc2015invariance,bruintjes2023what,cohen2015transformation,engstrom2019exploring}. Consistent with this finding, we show in Figure~\ref{fig_realdata_largemodel} that representations of rotated versions of the same image trace out non-trivial manifolds, and that their RSMs vary systematically with relative rotation angle, as measured by CKA. This experiment does not show that these representations are functionally equivalent in the same sense as the autoencoders considered above, but it illustrates that standard general-purpose vision models need not render latent stimulus rotations invisible to RSM-based model comparisons. Indeed, it is quite likely that the object recognition tasks on which these models were trained do not respect exact rotation invariance \citep{bruintjes2023what}. Thus, we emphasize that this analysis is intended to be an existence proof that these pretrained models can have gauge-dependent RSMs, not an attempt to establish the general prevalence, magnitude, or functional (in)significance of that gauge-dependence.

\section{Discussion}

Representational similarity measures have been used extensively in neuroscience and ML to compare internal representations of artificial and biological networks \citep{kriegeskorte2008representational,lampinen2025biases,zv2025summary,han2023identification,sucholutsky2025getting}. In this work, we demonstrate in simple settings that such measures may not be invariant under transformations that respect symmetries in the data. Specifically, we showed that functionally equivalent arrangements of receptive fields related by a gauge transformation can lead to qualitatively different RSMs. 

We refer to the gauge-dependence of the RSM as a confound because it can affect the outcome of an RSM-based analysis, and thus must be considered when interpreting such results. We emphasize that we do not intend to claim that it is always a bug. As we have seen, gauge transformations can relate fully-mixed codes to axis-aligned ones (Figure~\ref{fig1}); distinguishing them may be useful if the scientific question at hand is interested in these properties. Conversely, the gauge is functionally irrelevant for tasks that respect the symmetry. This equivalence is far stronger than equality of reconstruction error alone: representations related by a gauge transformation are isomorphic under a measure-preserving relabeling of the stimulus space, even under standard models of neural noise. Thus, our goal is not to argue against the use of the RSM, but to point out that differences in representational geometry reported by the RSM need not imply differences in symmetry-respecting function. 

As mentioned in the Introduction, we are not the first to observe that functionally-irrelevant transformations of a representation can change the RSM, nor the first to point out that its interpretation is subject to confounds. RSM-based analyses like CKA and RSA are confounded by finite-sample estimation bias \citep{song2012feature,chun2025estimating,pospisil2024estimating,murphy2024correcting}, by dataset and optimization details \citep{ciernik2025objective}, and by structure inherited from the probe inputs \citep{cui2022deconfounded,DujmoviBiorxiv}.\footnote{We note that \citet{cui2022deconfounded}'s proposed approach for de-confounding stimulus structure effects cannot in general remove gauge-dependence; see Appendix~\ref{app:cui}.} Many of these works focus on showing that highly dissimilar systems may appear, under various measures, to have similar RSMs \citep{murphy2024correcting,cui2022deconfounded,DujmoviBiorxiv}. Our work focuses on the converse: stimulus symmetries lead to functionally-similar systems with very different RSMs. Moreover, a range of previous works have constructed representational transformations that preserve specific downstream decoding functions while disrupting the RSM \citep{davari2023reliability,ding2021grounding,braun2025not}. Relative to this past literature, we are, to our knowledge, the first to specifically study how the restricted transformations induced by stimulus symmetries change the RSM. 

Theoretical work on the statistical physics of neural networks suggests that various summary statistics---including RSMs---may be stable when the number of neurons in each hidden layer of the network tends to infinity \citep{zv2025summary}. Our results on when the RSM variability tends to zero with the number of neurons in \S\ref{sec:factors} are compatible with those works. However, we showed that there are regimes in which learning may lead to solutions with a low \textit{effective} number of neurons. In evaluating how high the effective neuron count must be in order to suppress variability in the RSM, one must consider the intrinsic dimensionality of the symmetric stimulus manifold. From the curse of dimensionality, we expect the number of neurons required to suppress variability in the RSM should scale exponentially with the intrinsic dimension of the manifold. 

Ongoing drift in representations despite stable task performance has been observed in the brain and in artificial neural networks \citep{rule2019causes,masset2022drifting}. Previous works have sought to explain this phenomenon based on internal symmetries, yielding models with changing representations but stable RSMs \citep{qin2023coordinated, pashakhanloo2023drift,pashakhanloo2025contribution,masset2022drifting}. Complementing these results, our work shows how extrinsic stimulus symmetries can also lead to drifting representations over learning, with RSMs that change over time. Note that changing RSMs can occur even in networks that explicitly or implicitly impose a loss to align the RSMs of the stimuli and the representations (\textit{e.g.}, similarity matching networks as in \citep{qin2023coordinated,sengupta2018tiling,pashakhanloo2025contribution}). This is because symmetry along the latent dimension can allow RSMs to vary, while keeping the similarity matching objective fixed (see Fig.~\ref{fig_simmatching_supp} and App.~\ref{app:simmatch}). Additionally, there is experimental evidence showing changing RSMs over time in cortex \citep{schoonover2021representational,masset2022drifting}. Our results here show that, to reliably interpret the functional significance of such observations, one account for (potentially unknown) symmetries in data. 

A growing body of work aims to characterize the conditions under which networks have unique optimal representations. From a neuroscience perspective, a key question is when biologically-inspired constraints---like non-negativity of neural activity or energy constraints that can be modeled by regularization---select a unique RSM \citep{kunin2019loss,whittington2023disentanglement,dorrell2025range,braun2025not,brown2023privileged,khosla2024privileged,bordelon2022population}. These works, more broadly, seek to determine when axis-aligned solutions, in which each hidden neuron in an autoencoder responds only to a single latent dimension, are optimal \citep{whittington2023disentanglement,dorrell2025range}. However, how symmetries in data affect these results remains unknown. Indeed, in the presence of symmetries, there may be no preferred basis for the latent stimulus, so the very notion of axis-alignment can become ill-defined. Here, we showed that in the presence of continuous data symmetries, the constraints of energy minimization (\textit{e.g.} in the form of an $L_1$ penalty on neural activity) and non-negativity may not be sufficient to select a unique RSM. Precisely how little symmetry is required to remove a preference for axis alignment under a given objective will be an interesting question for future work.

In the representations we study, the underlying reason why the RSM is not gauge-invariant is that solutions with different gauge angle are not related by a simple rotation in representation space (Fig.~\ref{fig_man_schem}). We have studied manifold-tiling neural codes as a test case for this phenomenon, as they provide a minimal, neuroscience-inspired example of a nonlinear neural code. However, the idea that data symmetries can lead one to erroneously recognize two functionally equivalent representations as being distinct generalizes beyond the manifold-tiling setting, because many representations are not constrained to be orthogonally equivariant. To compare representations in a way that is invariant to data symmetry, one must design a metric that is invariant to the symmetry transformations, which in general requires knowledge of the stimulus space. Recent work on comparing representations using the pullback metric on the intrinsic data manifold could potentially address this problem \citep{zv2025how,caycogajic2026geometry}, but such methods require complete knowledge of the stimulus distribution. Absent such strong priors, it is unclear how to account for the gauge-invariance of representations. Our study motivates future attempts at seeking such metrics.

\clearpage

\section*{Acknowledgements}
We thank Juan Carlos Fern{\'a}ndez del Castillo and William Qian for helpful comments on a previous version of this manuscript. F.P. was supported by the Harvard Center for Brain Science (CBS)-NTT Fellowship Program on the Physics of Intelligence. J.A.Z.-V. and this research were supported by the Office of the Director of the NIH under Award Number DP5OD037354, and additionally by a Junior Fellowship from the Harvard Society of Fellows.

\section*{Code Availability}
The codes and functions to reproduce the main results of this paper are publicly available at: \href{https://github.com/fpashakhanloo/stimulus-sym-rsa}{https://github.com/fpashakhanloo/stimulus-sym-rsa}.

\section*{Declaration of LLM Usage}

The development of this research did not involve LLMs in its core aspects. The only LLM-based result is Figure~\ref{fig_realdata_largemodel}; we used OpenAI's Codex to help generate code for the analyses and plots reported there. We subsequently checked the generated code and plots.

\bibliographystyle{tmlr}
\bibliography{references}


\clearpage 
\appendix

\setcounter{figure}{0} 
\renewcommand{\thefigure}{S\arabic{figure}} 
\renewcommand{\theHfigure}{S\arabic{figure}}

\tableofcontents

\addtocontents{toc}{\protect\setcounter{tocdepth}{2}}

\clearpage 

\section{Additional results for toy model with \texorpdfstring{$n=4$}{n=4} neurons}

\subsection{Readout} \label{readout}
In our two-layer toy model, the gauge symmetry rotated the first layer weights. For completeness, here we show that a similar transformation applied to the readout weights leaves the input-to-output mapping and hence the loss unchanged. If $W \in \mathbb{R}^{n \times 2}$ and $U \in \mathbb{R}^{2 \times n}$ are the first and second layer weights of the network, it is easy to check that for $n=4$, the solution satisfies: $U = W^T$.
This essentially means that the outgoing weights of each neuron, $u_i$, tracks the incoming weights $w_i$, to have: $u_i = [\cos(\theta_i),\sin(\theta_i)]$ for $\theta_{i} = 2(i-1)\pi/n + \varphi$. To verify this, take the arbitrary stimulus $s_{\beta}=[\sin(\beta),\cos(\beta)]^T$, where $-\varphi \leq \beta < \pi/2-\varphi$. The output for this stimulus is:
\begin{align}
    y_{\beta} = U h_{\beta} &= 
    \begin{pmatrix}
        \cos(\theta_1)&\cos(\theta_2)&\cos(\theta_3)&\cos(\theta_4) \\\sin(\theta_1)&\sin(\theta_2)&\sin(\theta_3)&\sin(\theta_4)
    \end{pmatrix}\begin{pmatrix}
\sin(\varphi + \beta) \\
\cos(\varphi + \beta) \\
0 \\
0
\end{pmatrix} \\ &=
\begin{pmatrix}
\cos(\varphi)\sin(\varphi+\beta) - \sin(\varphi)\cos(\varphi+\beta) \\
\sin(\varphi)\sin(\varphi+\beta) + \cos(\varphi)\cos(\varphi+\beta)  
\end{pmatrix} \\&= 
\begin{pmatrix}
    \sin(\beta) \\ \cos(\beta)
\end{pmatrix} \\ &= s_{\beta}.
\end{align}
By symmetry, this shows that gauge transformations do not change the network function for any stimulus or gauge angle. See also Appendix \ref{app:sphere_reconstruction} for a more general analysis of reconstruction that covers this special case. 

\subsection{RSM for arbitrary angles} \label{arbitraryanglen=4}
Previously, we considered orthogonal stimuli $s_1=[1,0]^T$, $s_2=[0,1]^T$, etc. for calculating RSM. Here, we calculate RSMs for arbitrary angles. Consider the following stimuli:
\begin{align}
    s_{\alpha}=[\cos(\alpha),\sin(\alpha)]^T, \quad s_{\beta}=[\sin(\beta),\cos(\beta)]^T, \quad 
\end{align}
for $-\pi/2 + \varphi \leq \alpha \leq \varphi$  and $-\varphi \leq \beta <\pi/2 - \varphi$. Here, $\alpha$ and $\beta$ denote deviations from $s_1$ and $s_2$ respectively, and by symmetry this covers any arbitrary angle on the circle. The representations of stimuli $s_1$, $s_{\alpha}$ and $s_{\beta}$ are as follows:
\begin{align}
h_1 = \begin{pmatrix}
\cos(\varphi) \\
0 \\
0 \\
\sin(\varphi)
\end{pmatrix}, \quad
h_{\alpha} = \begin{pmatrix}
\cos(\varphi - \alpha) \\
0 \\
0 \\
\sin(\varphi-\alpha)
\end{pmatrix}.
\quad
h_{\beta} = \begin{pmatrix}
\sin(\varphi + \beta) \\
\cos(\varphi + \beta) \\
0 \\
0
\end{pmatrix}.
\end{align}
\begin{align}
\end{align}
This leads to the following RSM entries:
\begin{align}
RSM_{s_1,s_{\alpha}}&=\cos(\varphi)\cos(\varphi-\alpha) + \sin(\varphi)\sin(\varphi-\alpha) = \cos(\alpha) \\
RSM_{s_1,s_{\beta}}&=\cos(\varphi)\sin(\varphi+\beta)
\end{align}
The first RSM does not depend on the gauge variable. This is a case where the set of activated neurons between two stimuli are identical. The second RSM depends on $\varphi$. Note that, in general, even if two stimuli are nearby, there could be a range of $\varphi$ for which the set of activated neurons are different. This leads to a normalized variability ($\Delta$) that varies as function of angles between two stimuli. See Fig~\ref{fig_stim_supp}b for such a curve. For the generalization of these results to many neurons, see Appendix \ref{app:arbitrary_angle_many_neurons}. 

\subsection{Action of the gauge group on the four-neuron representation of the circle}

One might ask what the concrete form of the action of the $\mathrm{SO}(2)$ gauge group on the four-neuron representation of the circle is. Consider the representation of an arbitrary angle $\theta$:
\begin{align}
    h(\theta) = \begin{pmatrix} \ReLU(\cos \theta) \\ \ReLU(\sin\theta) \\ \ReLU(-\cos\theta) \\ \ReLU(-\sin\theta)\end{pmatrix}
\end{align}
The gauge group acts via
\begin{align}
    g_{\varphi} \cdot h( \theta) = h(\theta - \varphi). 
\end{align}
It is clear that this is a piecewise-linear action on the representation, and one can see that there are 16 distinct linear regions corresponding to distinct activation patterns before and after the gauge transformation. For instance, suppose that we are in the region
\begin{align}
    \theta \in [0, \pi/2], \varphi \in [0,\theta) \cup [3\pi/2+\theta, 2\pi)
\end{align}
such that
\begin{align}
    h(\theta) = \begin{pmatrix} \cos\theta \\ \sin\theta \\ 0 \\ 0 \end{pmatrix} \quad \textrm{and} \quad h(\theta-\varphi) = \begin{pmatrix} \cos(\theta-\varphi) \\ \sin(\theta-\varphi) \\ 0 \\ 0 \end{pmatrix} . 
\end{align}
Then, we have 
\begin{align}
    g_{\varphi} \cdot h(\theta) = \begin{pmatrix} \cos(\varphi) & \sin(\varphi) & 0 & 0 \\ -\sin(\varphi) & \cos(\varphi) & 0 & 0 \\ 0 & 0 & 0 & 0 \\ 0 & 0 & 0 & 0 \end{pmatrix} h(\theta). 
\end{align}
If instead we have that 
\begin{align}
    \theta \in [0, \pi/2) , \varphi \in [\theta,\theta+\pi/2),
\end{align}
such that 
\begin{align}
    h(\theta) = \begin{pmatrix} \cos\theta \\ \sin\theta \\ 0 \\ 0 \end{pmatrix} \quad \textrm{and} \quad h(\theta-\varphi) = \begin{pmatrix} \cos(\theta-\varphi)  \\ 0 \\ 0 \\ -\sin(\theta-\varphi)\end{pmatrix} , 
\end{align}
then
\begin{align}
    g_{\varphi} \cdot h(\theta) = \begin{pmatrix} \cos(\varphi) & \sin(\varphi) & 0 & 0  \\ 0 & 0 & 0 & 0 \\ 0 & 0 & 0 & 0 \\ \sin(\varphi) & -\cos(\varphi) & 0 & 0 \end{pmatrix} h(\theta).
\end{align}
One could enumerate the matrices for each linear region, but we will not do so. 

We note that this shows that the representation $h(\theta)$ is not in fact a linear representation in the group-theoretic sense, as the group action is not linear. 

\subsection{Alignment between two RSMs}\label{app:4neuroncka}

Consider two instances of the four-neuron representation, with distinct gauge angles $\varphi_{1}$ and $\varphi_{2}$. Using the form of the RSM from \eqref{RSMtoy}, we can work out that the raw kernel alignment between their RSMs is
\begin{align}
    \operatorname{KA}(\varphi_{1},\varphi_{2}) 
    &= \frac{\tr(RSM_{\varphi_{1}} RSM_{\varphi_{2}})}{\sqrt{\tr(RSM_{\varphi_{1}} RSM_{\varphi_{1}})\tr(RSM_{\varphi_{2}} RSM_{\varphi_{2}})}}
    \\
    &= \frac{1 + 2 \rho(\varphi_{1})\rho(\varphi_{2})}{\sqrt{(1+2\rho(\varphi_{1})^2)(1+2\rho(\varphi_{2})^2)}}
\end{align}
where $\rho(\varphi) = \sin\varphi \cos\varphi$ with $\varphi$ taken modulo $\pi/2$. Defining the $4 \times 4$ centering matrix $C$ by $C_{ij} = \delta_{ij} - 1/4$, the centered kernel alignment \citep{williams2024equivalence} is
\begin{align}
    \operatorname{CKA}(\varphi_{1},\varphi_{2}) 
    &= \frac{\tr(C RSM_{\varphi_{1}} C RSM_{\varphi_{2}})}{\sqrt{\tr(C RSM_{\varphi_{1}} C RSM_{\varphi_{1}})\tr(C RSM_{\varphi_{2}} C RSM_{\varphi_{2}})}}
    \\
    &= \frac{3 - 2 (\rho(\varphi_{1})+\rho(\varphi_{2})) + 4 \rho(\varphi_{1}) \rho(\varphi_{2})}{\sqrt{(3- 4 \rho(\varphi_{1}) + 4 \rho(\varphi_{1})^2)(3-4\rho(\varphi_{2})+4\rho(\varphi_{2})^2)}}. 
\end{align}
Both of these expressions are gauge-dependent. Because of the properties of $\rho(\varphi)$, they are doubly-periodic, and both are minimized when $\rho(\varphi_{1})$ and $\rho(\varphi_{2})$ are maximally distinct. This occurs when one of $\varphi_{1}$ or $\varphi_{2}$ is zero modulo $\pi/2$, while the other is $\pi/4$ modulo $\pi/2$, which leads to either $\rho(\varphi_{1})$ or $\rho(\varphi_{2})$ vanishing while the other is equal to $1/2$. In such a case both the raw KA and the CKA are equal to $\sqrt{2/3} \approx 0.82$. We plot the resulting matrices of centered and non-centered kernel alignment in Figure~\ref{fig_cka_supp}.

\section{Deferred derivations related to general formalism}\label{app:formal}

\subsection{Gauge-invariance of performance on invariant tasks}\label{app:decoding_proof}

Here, we prove that invariant tasks (those satisfying $t(g\cdot x) = t(x)$) have gauge-invariant joint distributions $(u,t(x))$ with the representation. To see this, fix measurable subsets $A \subseteq H$ and $B \subseteq Y$, and consider
\begin{align}
    \mathbb{P}_{g \cdot \rho}[ (u,t(x)) \in A \times B ] 
    &= \int_{X} (g\cdot\rho)(A \mid x) \mathbf{1}_{t(x) \in B}\, \mu(dx)
    \\
    &= \int_{X} \rho(A \mid g^{-1} \cdot x) \mathbf{1}_{t(x) \in B}\, \mu(dx) && (\textrm{def. of $g \cdot \rho$})
    \\
    &= \int_{X} \rho(A \mid x') \mathbf{1}_{t(g\cdot x') \in B}\, \mu(dx') && (x' = g^{-1} \cdot x \textrm{ and $G$-inv. of $\mu$})
    \\
    &= \int_{X} \rho(A \mid x') \mathbf{1}_{t(x') \in B}\, \mu(dx') && (\textrm{$G$-inv. of $t$})
    \\
    &=  \mathbb{P}_{\rho}[ (u,t(x)) \in A \times B ], 
\end{align}
which proves the claim.

\subsection{Conditions for gauge-invariance of the RSM}

Here, we prove the claim of Section~\ref{sec:rsmgauge}, which gives the conditions under which the RSM is gauge-invariant. Recall that our goal is to show that, for some fixed encoder $h$, 
\begin{align}
    RSM_{g\cdot h}(x,x') = RSM_{h}(x,x')
\end{align}
for any two inputs $x,x'\in X$ and any group element $g \in G$ if and only if there exists an orthogonal matrix $O(g) \in \mathcal{O}(n)$, depending on $g$ but not $x$, such that
\begin{align} \label{eqn:orthsupp}
    (g \cdot h)(x) = O(g) h(x)
\end{align}
for all $x \in X$. The matrix $O(g)$ is not constrained outside the subspace $V = \linspan\{h(x)\,:\,x \in X\}$ spanned by the representations. However, this non-uniqueness is relevant neither for the RSM nor for readout functionality.

It is immediately clear that it is sufficient for \eqref{eqn:orthsupp} to be satisfied for the RSM to be gauge-invariant, as under \eqref{eqn:orthsupp} we have
\begin{align}
    RSM_{g\cdot h}(x,x') = h(x)^{\top} O(g)^{\top} O(g) h(x') = h(x)^{\top} h(x') = RSM_{h}(x,x'). 
\end{align}

What remains is to show that this condition is necessary. Fix a group element $g \in G$. Writing
\begin{align}
    \tilde{h}(x) = (g\cdot h)(x),
\end{align}
we assume that the RSM is gauge-invariant, so that
\begin{align}
    \tilde{h}(x)^{\top} \tilde{h}(x') = h(x)^{\top} h(x')
\end{align}
for all $x,x'\in X$. Let  
\begin{align}
    V = \linspan\{h(x)\,:\,x\in X\} \subseteq \mathbb{R}^{n}
\end{align}
be the linear subspace of $\mathbb{R}^{n}$ spanned by the representation. Because $V$ is finite-dimensional, we may choose a finite set of $p \leq n$ inputs $\{x_{1},\ldots,x_{p}\}$ such that $\{h(x_{1}),\ldots,h(x_{p})\}$ is a basis for $V$. Consider the gauge-transformed representations $\{\tilde{h}(x_{1}),\ldots,\tilde{h}(x_{p})\}$. By assumption, they have the same Gram matrix as the basis $\{h(x_{1}),\ldots,h(x_{p})\}$. It is a classical fact (see Theorem 7.3.11 of \citet{horn2012matrix}) that two real vector realizations of the same Gram matrix are related by an orthogonal transformation, \textit{i.e.}, there must exist a matrix $O = O(g) \in \mathcal{O}(n)$ such that
\begin{align}
    \tilde{h}(x_i) = O(g) h(x_i), \quad i \in [p],
\end{align}
where we emphasize that the matrix is allowed to depend on $g$. What remains is to show that $O h(x) = \tilde{h}(x)$ for all $x \in X$. Fixing an input $x \in X$, we therefore want to show that 
\begin{align}
    \Vert \tilde{h}(x) - O h(x) \Vert^2 = \tilde{h}(x)^{\top} \tilde{h}(x) - 2 \tilde{h}(x)^{\top} O h(x) + h(x)^{\top} O^{\top} O h(x)
\end{align}
vanishes. By the assumption of RSM invariance, $\tilde{h}(x)^{\top} \tilde{h}(x) = h(x)^{\top} h(x)$, and by the fact that $O$ is orthogonal, $h(x)^{\top} O^{\top} O h(x) = h(x)^{\top} h(x)$. This leaves us with the cross term. Because $h(x) \in V$, we can expand it in the basis $\{h(x_1),\ldots,h(x_{p})\}$ as
\begin{align}
    h(x) = \sum_{i=1}^{p} a_{i} h(x_{i})
\end{align}
for some set of coefficients $a_{1},\ldots,a_{p}$. Then, the cross term expands as
\begin{align}
    \tilde{h}(x)^{\top} O h(x)
    &= \sum_{i=1}^{p} a_{i} \tilde{h}(x)^{\top} O h(x_i)
    \\
    &= \sum_{i=1}^{p} a_{i} \tilde{h}(x)^{\top} \tilde{h}(x_i) && (\tilde{h}(x_i) = O h(x_i))
    \\
    &= \sum_{i=1}^{p} a_{i} h(x)^{\top} h(x_i) && (\textrm{RSM invariance})
    \\
    &= h(x)^{\top} h(x) && (\textrm{basis expansion of } h(x)).
\end{align}
Therefore, we have
\begin{align}
    \Vert \tilde{h}(x) - O h(x) \Vert^2 &= \tilde{h}(x)^{\top} \tilde{h}(x) - 2 \tilde{h}(x)^{\top} O h(x) + h(x)^{\top} O^{\top} O h(x)
    \\
    &= h(x)^{\top} h(x) - 2 h(x)^{\top} h(x) + h(x)^{\top} h(x)
    \\
    &= 0, 
\end{align}
which proves that $\tilde{h}(x) = O(g) h(x)$ for the arbitrary input $x \in X$, and thus for all inputs. Importantly, the above argument shows that the matrix $O$ is determined only on the span $V$ of the representation, not on the orthogonal subspace $V^{\perp}$.  We therefore conclude the desired claim. 

In the analysis above, our starting assumption was that the full RSM was invariant for any pair of stimuli $x,x' \in X$. If we instead observed equality of the finite RSMs resulting from a set of trial stimuli, we could not guarantee that \eqref{eqn:orthsupp} holds globally. If the trial stimuli form a spanning set in the sense that they span $V$, then one could reconstruct the candidate matrix $O$, but showing that this extends to all un-tested stimuli requires stronger assumptions. In the proof above, this was possible thanks to the assumption of invariance of the RSM as a kernel function. 

An alternative sufficient condition would be that we can write the full representation as an interpolant between the representations of the trial stimuli, \textit{i.e.}, if there exist weights $\omega_{i} : X \to \mathbb{R}$ such that
\begin{align}
    h(x) = \sum_{i=1}^{p} \omega_{i}(x) h(x_{i})
\end{align}
for any $x \in X$, where $\{x_{1},\ldots,x_{p}\}$ is the fixed set of trial stimuli on which we know the RSM is invariant. These weight functions must satisfy
\begin{align}
    \omega_{i}(x_{j}) = \delta_{ij}
\end{align}
for any probe stimulus, and crucially must be shared by the original and transformed representations, in the sense that 
\begin{align}
    \tilde{h}(x) = \sum_{i=1}^{p} \omega_{i}(x) \tilde{h}(x_{i})
\end{align}
for any $x \in X$. Then, supposing that the RSMs for the trial stimuli are invariant, we know that there is an $O$ such that $\tilde{h}(x_{i}) = O h(x_{i})$ for all $i \in [p]$, and thus
\begin{align}
    \tilde{h}(x) = \sum_{i=1}^{p} \omega_{i}(x) \tilde{h}(x_{i}) = \sum_{i=1}^{p} \omega_{i}(x) O h(x_{i}) = O \sum_{i=1}^{p} \omega_{i}(x) h(x_{i}) = O h(x) 
\end{align}
for any $x \in X$. 

\subsection{Similarity matching between the stimulus and the representation}\label{app:simmatch}

A natural question is whether objectives that explicitly aim to match the RSM to the stimulus similarity can fix a gauge. The simplest among such objective functions would be similarity matching \citep{sengupta2018tiling,qin2023coordinated,pashakhanloo2025contribution}, which would seek to minimize a population loss of the general form
\begin{align}
    \mathcal{C}[h] = \mathbb{E}_{z,z'\sim\mu_{Z},\xi,\xi'\sim\mu_{\Xi}}[ (k( z,z' ) - h(x)^{\top} h(x') )^2 ]
\end{align}
where $x = x(z,\xi)$ and $x' = x(z',\xi')$. Here, we assume that the latent space is equipped with a $G$-invariant similarity function $k$, \textit{i.e.}, $k(\cdot,\cdot)$ is a symmetric function that satisfies
\begin{align}
    k( g \cdot z, g \cdot z' ) = k(z,z')
\end{align}
for any $g \in G$ and any $z,z' \in Z$. We claim that such a loss is gauge-invariant, in the sense that
\begin{align}
    \mathcal{C}[g \cdot h] = \mathcal{C}[h]
\end{align}
for any $g \in G$. The proof is straightforward:
\begin{align}
    \mathcal{C}[g \cdot h] 
    &= \mathbb{E}[ (k(z,z') - (g\cdot h)(x(z,\xi))^{\top} (g\cdot h)(x(z',\xi')) )^2 ]
    \\
    &= \mathbb{E}[ (k(z,z') - h(x(g^{-1} \cdot z,\xi))^{\top} h(x(g^{-1} \cdot z',\xi')) )^2 ] && (\textrm{def. of } g \cdot h)
    \\
    &= \mathbb{E}[ (k( g \cdot z, g\cdot z') - h(x(z,\xi))^{\top} h(x(z',\xi')) )^2 ] && (\textrm{$G$-inv. of } \mu_{Z})
    \\
    &= \mathcal{C}[h] && (\textrm{$G$-inv. of } k). 
\end{align}
Now, in practice, one minimizes an empirical loss for a finite number of samples; this could in principle break the symmetry. However, as illustrated in Figure~\ref{fig_simmatching_supp}, even with a relatively modest number of examples the similarity matching loss can remain nearly gauge-independent.

\section{Detailed analysis of the toy orientation tuning model with many neurons} \label{AppendixA}

In the main text, we focused on the case where the toy orientation tuning model has only $n=4$ neurons. This is the minimum number of neurons required to faithfully reconstruct the ring because (1) at least one neuron must respond to every stimulus, and (2) if only a single neuron responds to some set of stimuli, then responses may be ambiguous due to the symmetry in each neuron's response under reflection about the axis of its preferred stimulus. This, for instance, means that $n=3$ neurons does not suffice, as with equidistributed receptive field centers there are thus regions around each neuron's receptive field center where only a single neuron is active. In this Appendix, we consider $n>4$ neurons. We show that the RSM for the toy orientation tuning model has a simple form whenever the number of neurons is an integer multiple of four. 

As in the main text, the activation of the $i$-th of $n$ neurons is given by $h_{i}(x) = \ReLU(w_{i}^{\top} x)$, where $w_{i} = (\cos\theta_i, \sin\theta_{i})^{\top}$ for $\theta_{i} = 2(i-1)\pi/n + \varphi$. The four test stimuli of interest are $s_{1} = (1,0)^{\top}$, $s_{2} = (0,1)^{\top}$, $s_{3} = (-1,0)^{\top}$, and $s_{4} = (0,-1)^{\top}$, and the RSM has elements $RSM_{ab} = \sum_{i=1}^{n} h_{i}(s_{a}) h_{i}(s_b)$ for $a,b \in [4]$. 

\subsection{Reconstruction}\label{sec:many_neuron_reconstruction}

Before considering the RSM, we show that reconstruction is gauge-invariant for any even $n$. Let $n=2m$. Then, for any $\varphi$, we can write the weight matrix as 
\begin{align}
    W &= 
    \begin{pmatrix}
        \cos( \varphi ) & \sin(\varphi) 
        \\ 
        \cos(\pi/m + \varphi) & \sin(\pi/m + \varphi) \\ 
        \vdots 
        \\
        \cos(\pi (m-1)/m + \varphi ) & \sin(\pi (m-1)/m + \varphi )
        \\
        \cos(\pi+ \varphi ) & \sin(\pi + \varphi )
        \\
        \cos(\pi + \pi/m + \varphi ) & \sin(\pi + \pi/m + \varphi )
        \\
        \vdots 
        \\
        \cos(\pi + \pi (m-1)/m +\varphi) & \sin(\pi + \pi (m-1)/m + \varphi) 
    \end{pmatrix}
    = \sqrt{\frac{m}{2}} \begin{pmatrix} U \\ - U \end{pmatrix}, 
\end{align}
where
\begin{align} 
    U = \sqrt{\frac{2}{m}}\begin{pmatrix}
        \cos( \varphi ) & \sin(\varphi) 
        \\ 
        \cos(\pi/m + \varphi) & \sin(\pi/m + \varphi) \\ 
        \vdots 
        \\
        \cos(\pi (m-1)/m + \varphi ) & \sin(\pi (m-1)/m + \varphi )
    \end{pmatrix} \in \mathbb{R}^{m \times 2}. 
\end{align}
The matrix $U$ has the important property that it is semi-orthogonal: 
\begin{align}
    U^{\top} U = I_{2},
\end{align}
as can be verified by computing each of its elements:
\begin{align}
    (U^{\top} U)_{11} &= \frac{2}{m} \sum_{k=0}^{m-1} \cos( \pi k/m + \varphi)^{2} = 1
    \\
    (U^{\top} U)_{12} = (U^{\top} U)_{21} &= \frac{2}{m} \sum_{k=0}^{m-1} \cos( \pi k/m + \varphi) \sin( \pi k/m + \varphi) = 0
    \\
    (U^{\top} U)_{22} &= \frac{2}{m} \sum_{k=0}^{m-1} \sin( \pi k/m + \varphi)^{2} = 1 . 
\end{align}
Using the general result we prove in Appendix \ref{app:sphere_reconstruction} for reconstructions of the $(d-1)$-sphere $\mathbb{S}^{d-1}$ in the special case $d=2$, we therefore conclude that perfect decoding for any gauge angle is possible by using $\frac{2}{m} W^{\top}$ as the decoding weights, where the normalization factor does not matter because the ReLU is positive-homogeneous. In other words, for any gauge angle $\varphi$ we have
\begin{align}
    \frac{2}{m} W^{\top} \ReLU(W x) = x 
\end{align}
for all $x \in \mathbb{S}^{1}$.

\subsection{Evaluation of the similarity matrix elements}

By symmetry, all diagonal elements of the RSM are equal to:
\begin{align}
    RSM_{diag} = \sum_{k=0}^{n-1} \ReLU\left(\cos\left(\frac{2\pi k}{n} + \varphi \right)\right)^2,
\end{align}
and non-zero off-diagonal elements are all equal to:
\begin{align}
    \rho = \sum_{k=0}^{n-1} \ReLU\left(\cos\left(\frac{2\pi k}{n} + \varphi \right)\right) \ReLU\left(\sin\left(\frac{2\pi k}{n} + \varphi \right)\right) .
\end{align}

We can see that increasing $\varphi$ by integer multiples of $2\pi/n$ is equivalent to shifting the index $k$, which has the effect of circularly permuting neuron labels. As the RSM is invariant under permutation of the neuron labels, it must therefore be periodic in $\varphi$, with period $2\pi/n$. Therefore, it suffices to consider $\varphi \in [0, 2\pi/n)$. We therefore put $\varphi = 2\pi \delta/n$ for $\delta \in [0,1)$.

First, consider the off-diagonal elements. As $\cos(\theta)$ and $\sin(\theta)$ are simultaneously positive only for $\theta \in [0,\pi/2)$,
\begin{align}
    \rho = \sum_{k=0}^{n-1} \cos\left(\frac{2\pi (k+\delta)}{n}\right) \sin\left(\frac{2\pi (k+\delta)}{n} \right) \mathbf{1}_{0 \leq k+\delta < n/4}
\end{align}
Supposing that $n =  4q$ for an integer $q\geq 1$, we see that $0 \leq k + \delta < n/4$ only if $k \leq q-1$, for any $\delta \in [0,1)$. Therefore, 
\begin{align}
    \rho = \sum_{k=0}^{q-1} \cos\left(\frac{\pi (k+\delta)}{2q}\right) \sin\left(\frac{\pi (k+\delta)}{2q} \right) 
\end{align}
This is at last a sum that can be evaluated, giving
\begin{align}
    \rho = \frac{1}{2} \cot\left(\frac{\pi}{2q}\right) \cos\left( \frac{\pi\delta}{q} \right) + \frac{1}{2} \sin\left(\frac{\pi\delta}{q} \right)
\end{align}
or, in terms of $n$ and $\varphi$,
\begin{align}
    \rho = \frac{1}{2} \cot\left(\frac{2\pi}{n}\right)\cos(2\varphi) + \frac{1}{2} \sin(2\varphi). 
\end{align}

As $\cos(\theta)$ is positive if $\theta \in [0,\pi/2) \cup (3\pi/2,2\pi)$, we can similarly evaluate the diagonal elements as 
\begin{align}
    \sum_{k=0}^{n-1} \cos\left(\frac{2\pi (k+\delta)}{n} \right)^2 [\mathbf{1}_{0 \leq k+\delta < q} + \mathbf{1}_{3 q < k+\delta < 4 q }]
    &= \sum_{k=0}^{q-1} \cos\left(\frac{\pi (k+\delta)}{2q} \right)^2 + \sum_{k=3q}^{4q-1} \cos\left(\frac{\pi (k+\delta)}{2q} \right)^2
    \\
    &= q. 
\end{align}

Thus, letting
\begin{align} \label{rhon}
    \rho_{n}(\varphi) = \frac{1}{2} \cot\left(\frac{2\pi}{n}\right)\cos(2\varphi) + \frac{1}{2} \sin(2\varphi),
\end{align}
we find that for any $n$ a multiple of four we have
\begin{align} \label{RSMn}
  RSM = 
  \begin{pmatrix}
    \frac{n}{4} & \rho_{n}(\varphi) & 0 & \rho_{n}(\varphi) \\
    \rho_{n}(\varphi) & \frac{n}{4} & \rho_{n}(\varphi) & 0 \\
    0 & \rho_{n}(\varphi) & \frac{n}{4} & \rho_{n}(\varphi) \\
    \rho_{n}(\varphi) & 0 & \rho_{n}(\varphi) & \frac{n}{4}
\end{pmatrix}
\end{align}
for any $\varphi \in [0,2\pi/n)$. For larger $\varphi$, we can apply this result using $\varphi \mod 2\pi/n$.

We can evaluate by a similar argument the population-averaged firing rate for each of the test stimuli:
\begin{align}
    \bar{r} &= \frac{1}{n} \sum_{k=0}^{n-1} \cos\left(\frac{2\pi (k+\delta)}{n} \right) [\mathbf{1}_{0 \leq k+\delta < q} + \mathbf{1}_{3 q < k+\delta < 4 q }]
    \\&= \frac{1}{n}\sum_{k=0}^{q-1} \cos\left(\frac{\pi (k+\delta)}{2q} \right) + \frac{1}{n}\sum_{k=3q}^{4q-1} \cos\left(\frac{\pi (k+\delta)}{2q} \right)
    \\
    &= \frac{1}{n}\csc\left(\frac{\pi}{4q}\right) \cos\left(\frac{\pi-2\pi\delta}{4q}\right)
    \\
    &= \frac{1}{n}\csc\left(\frac{\pi}{n}\right)\cos\left(\frac{\pi}{n}-\varphi\right),
\end{align}
where again in the last line we consider $\varphi \in [0,2\pi/n)$. 

\subsection{Large-\texorpdfstring{$n$}{n} limit}

We now want to study what happens when we take the number of neurons to be large. To do so, it is convenient to use the parameterization of the mean firing rate and RSM in terms of $q$ and $\delta$, as the period of these objects in $\delta$ is defined to be independent of $n$. First, the mean firing rate expands as
\begin{align}
    \bar{r} = \frac{1}{4q}\csc\left(\frac{\pi}{4q}\right) \cos\left(\frac{\pi-2\pi\delta}{4q}\right) = \frac{1}{\pi} - \frac{\pi (1-6\delta+6\delta^2)}{48q^2} + \mathcal{O}\left(\frac{1}{q^4}\right).
\end{align}
The normalized non-zero off-diagonal elements of the RSM are given by
\begin{align}
    \frac{1}{2q} \cot\left(\frac{\pi}{2q}\right) \cos\left( \frac{\pi\delta}{q} \right) + \frac{1}{2q} \sin\left(\frac{\pi\delta}{q} \right) 
    &= \frac{1}{\pi} - \frac{\pi (1-6\delta+6\delta^2)}{12 q^2}  + \mathcal{O}\left(\frac{1}{q^4}\right).
\end{align}
This shows that the mean firing rate and RSM are gauge-invariant only asymptotically, in the limit $n \to \infty$.

\subsection{Normalized variability}

As in the main text, we consider the normalized variability
\begin{align}
    \Delta_{n} &= \frac{\max_{\varphi} RSM_{off} - \min_{\varphi} RSM_{off}}{RSM_{diag}} .
\end{align}
Using the expression for $RSM_{off}$ from above, we see that it is maximized at $\delta = 1/2$, where it takes value 
\begin{align}
    \frac{1}{2} \cot\left(\frac{\pi}{2q}\right) \cos\left( \frac{\pi\delta}{q} \right) + \frac{1}{2} \sin\left(\frac{\pi\delta}{q} \right) \bigg|_{\delta = 1/2} = \frac{1}{2} \csc\left(\frac{\pi}{2q}\right)
\end{align}
and minimized at $\delta= 0$, where it takes value
\begin{align}
    \frac{1}{2} \cot\left(\frac{\pi}{2q}\right) \cos\left( \frac{\pi\delta}{q} \right) + \frac{1}{2} \sin\left(\frac{\pi\delta}{q} \right) \bigg|_{\delta = 0} = \frac{1}{2} \cot\left(\frac{\pi}{2q}\right) .
\end{align}
Using the identity $\csc(\theta)-\cot(\theta)=\tan(\theta/2)$, we thus have that
\begin{align}
    \Delta_{4q} = \frac{1}{2q} \tan\left(\frac{\pi}{4q}\right) ,
\end{align}
or 
\begin{align}
    \Delta_{n} = \frac{2}{n}\tan\left(\frac{\pi}{n}\right)
\end{align}
in terms of $n$. As $\tan(\theta) = \theta + \mathcal{O}(\theta^{3})$ as $\theta\downarrow 0$, we thus find that
\begin{align}
    \Delta_{n} = \frac{2\pi}{n^2} + \mathcal{O}\left(\frac{1}{n^4}\right).
\end{align}

\subsection{Penalizing the \texorpdfstring{$L_1$}{L1} norm of activations does not fix a gauge} \label{l1_gauge}
In the main text, we showed that one way of promoting RSM variability with a large number of neurons is the presence of energetic cost (such as the L1-penalty) on the activations. Here, we show that when data is uniformly distributed on the manifold, L1-penalty does not fix a gauge.
Observe that the response of the $i$-th neuron to a stimulus $x = (\cos\psi,\sin\psi)^{\top}$ is simply
\begin{align}
    h_{i}(x) = \ReLU\left( \cos\left(\frac{2\pi (i-1)}{n} + \varphi - \psi \right) \right), \quad i \in [n] .
\end{align}
Then, the $L_p$ norm of the hidden layer activations, assuming a uniform angular distribution, is
\begin{align}
    \Vert h \Vert_{p} = \left( \int_{0}^{2\pi} \frac{d\psi}{2\pi} \bar{r}_{n,p}(\varphi-\psi) \right)^{1/p},
\end{align}
where we let
\begin{align}
    \bar{r}_{n,p}(\varphi-\psi) = \sum_{k=0}^{n-1} \ReLU\left( \cos\left(\frac{2\pi k}{n} + \varphi - \psi \right) \right)^{p} .
\end{align}
Though we have assumed that everything is normalized, we remark that if we allowed radial variation in the stimulus, we could ignore it in this analysis. This is because the positive-homogeneity of the ReLU means that adding a variable radius will simply multiply the formula above by a constant independent of both $n$ and $\varphi$.

By an argument identical to our previous analysis, $\bar{r}_{n,p}(\varphi-\psi)$ must be $2\pi/n$-periodic in its argument. Therefore, by averaging over $\psi \in [0,2\pi)$ we are averaging over $n$ periods, which in turn means that $\Vert h \Vert_{p}$ must be independent of the gauge angle $\varphi$. Thus, penalizing the norm of the activations cannot fix a gauge. 

How do deviations of the stimulus distribution from a uniform distribution on the full circle affect this result? In particular, what happens if we consider a discrete set of equally-spaced points on the circle, in which case the rotation symmetry is replaced by a cyclic group. Recalling our results from before, we have for $n$ an integer multiple of four that
\begin{align}
    \bar{r}_{n,1}(\varphi-\psi) = \csc\left(\frac{\pi}{n}\right)\cos\left(\frac{\pi}{n}-(\varphi-\psi)\right),
\end{align}
where we should plug in the residue of $\varphi-\psi$ modulo $2\pi/n$, while 
\begin{align}
    \bar{r}_{n,2}(\varphi-\psi) = \frac{n}{4} . 
\end{align}
This means that for $n$ a multiple of 4 penalizing the $L_2$ norm of activations cannot fix a gauge for the trivial reason that it is stimulus-independent. In contrast, the $L_1$ norm is in principle sensitive to fluctuations in the distribution of stimuli. To see whether this can fix a gauge in the case where the stimuli are discrete and equidistributed around the circle, first suppose that $n=4$. Then, 
\begin{align}
    \bar{r}_{4,1}(\phi-\psi) = \cos(\varphi - \psi) + \sin(\varphi-\psi)
\end{align}
must be evaluated using the residue of $\varphi-\psi$ modulo $\pi/2$. One can then, for instance, evaluate the total $L_1$ norm in the case where there are four stimuli equidistributed about the circle, which leads to 
\begin{align}
    \Vert h \Vert_{1} = \frac{1}{4} \sum_{k=0}^{3} \bar{r}_{4,1}(\varphi-\pi k/2) = \cos\varphi + \sin\varphi
\end{align}
for $\varphi$ taken modulo $\pi/2$. This is clearly not gauge-invariant. However, it does not prefer a unique $\varphi$, only some discrete set. This generalizes to larger numbers of stimuli, where including a penalty leads to a quantized set of preferred gauge angles $\varphi$.

\subsection{The RSM for arbitrary angular differences}\label{app:arbitrary_angle_many_neurons}

We can extend the analysis above to cover arbitrary angular displacements. To cover the circle, we consider $s_{1}$ and
\begin{align}
    s_{\alpha}=[\cos(\alpha),\sin(\alpha)]^{\top}
\end{align}
for $-\pi\leq \alpha \leq \pi$. With $n$ neurons, the RSM between the representations of the vectors $s_1$ and $s_{\alpha}$ is
\begin{align}
    RSM_{s_1,s_{\alpha}}(\varphi) 
    &= \sum_{k=0}^{n-1} \ReLU\left( \cos\left(\frac{2\pi k}{n} + \varphi\right) \right) \ReLU\left(\cos\left(\frac{2\pi k}{n} + \varphi - \alpha \right) \right) . 
\end{align}
By a rationale identical to our previous analysis, this is $2\pi/n$-periodic in the gauge angle $\varphi$, so it suffices to consider $\varphi \in [0,2\pi/n)$. This holds for any $\alpha$.

One important feature is that $RSM_{s_1,s_{\alpha}}$ is in general an even function of $\alpha$ only in the limit $n \to \infty$, where the dependence on $\varphi$ drops out. The RSM is also even in the special case $\varphi = 0$, 
\begin{align}
    RSM_{s_1,s_{-\alpha}}(\varphi=0) = RSM_{s_1,s_{\alpha}}(\varphi=0). 
\end{align}
This follows from the fact that we can transform $k \mapsto -k$ and use the fact that $\ReLU(\cos(\cdot))$ is even. However, in general, we can see that
\begin{align}
    RSM_{s_{1},s_{-\alpha}}(\varphi) = RSM_{s_{1},s_{\alpha}}(\varphi+\alpha). 
\end{align}
The symmetry---or lack thereof---at finite $n$ is clear geometrically.  Symmetry is recovered, however, for $\alpha = 2 \pi r/n$ for some integer $r$, as then the gauge shift from $\varphi$ to $\varphi+\alpha$ is precisely an integer multiple of $2\pi/n$ and thus does not change the RSM. Our earlier results, which only considered $\alpha$ as an integer multiple of $\pi/2$, fall precisely into this case. 

It is useful to observe that, in the $n\to\infty$ limit where the dependence on $\varphi$ drops out, we have
\begin{align}
    \lim_{n \to \infty} \frac{1}{n} RSM_{s_{1},s_{\alpha}} &= \frac{1}{2\pi} \int_{0}^{2\pi} \ReLU(\cos(\theta))\ReLU(\cos(\theta-\alpha)) \, d\theta. 
\end{align}
This is clearly a $2\pi$-periodic function of $\alpha$, and moreover is even in $\alpha$ as we are integrating over a full period. It is then clear that it suffices to consider $\alpha \in [0,\pi]$. Then, using the fact that both cosines are positive only over the intervals $(\max(0,\alpha-\pi/2),\pi/2)$ and $(\min(2\pi,3\pi/2+\alpha),2\pi)$, the problem reduces to a pair of elementary trigonometric integrals, which we can evaluate to obtain 
\begin{align}
    \lim_{n \to \infty} \frac{1}{n} RSM_{s_{1},s_{\alpha}} = \frac{(\pi-\alpha) \cos(\alpha) + \sin(\alpha)}{4\pi} . 
\end{align}
Using the symmetry and periodicity properties of the function, this determines its value for any $\alpha$. Up to normalization, this is the familiar arc-cosine kernel from \citet{cho2009kernel}, as we would expect. 

We now return to the discrete computation, which is a straightforward but slightly tedious exercise in trigonometric sums. As before, we proceed by letting $n=4q$ for some positive integer $q$, and by letting $\varphi = 2\pi \delta/n = \pi \delta / (2q)$ for some $\delta \in [0,1)$. This leads to 
\begin{align}
    RSM_{s_1,s_{\alpha}} 
    &= \sum_{k=0}^{4q-1} \ReLU\left( \cos\left(\frac{\pi (k+\delta)}{2q} \right) \right) \ReLU\left(\cos\left(\frac{\pi (k+\delta)}{2q} - \alpha \right) \right) . 
\end{align}
As $\cos(\theta)$ is positive if $\theta \in [0,\pi/2) \cup (3\pi/2,2\pi)$, we can expand the summation based on when the first rectified cosine is non-zero as
\begin{align}
    RSM_{s_1,s_{\alpha}} 
    &= \sum_{k\in\{0,\ldots, q-1\} \cup \{3q,\ldots,4q-1\}} \cos\left(\frac{\pi (k+\delta)}{2q} \right) \ReLU\left(\cos\left(\frac{\pi (k+\delta)}{2q} - \alpha \right) \right) .
\end{align}
We now make a further change of variables, and write
\begin{align}
    \varphi-\alpha = \frac{\pi}{2q} (m+\varepsilon)
\end{align}
where $m$ is an integer and $\varepsilon \in [0,1)$, in terms of which we have 
\begin{align}
    RSM_{s_1,s_{\alpha}} 
    &= \sum_{k\in\{0,\ldots, q-1\} \cup \{3q,\ldots,4q-1\}} \cos\left(\frac{\pi (k+\delta)}{2q} \right) \ReLU\left(\cos\left(\frac{\pi (k+m+\varepsilon)}{2q} \right) \right). 
\end{align}
As this function is $4q$-periodic in $m$, it is sufficient to consider the range $-2q \leq m \leq 2q$. 

Consider the first range in the summation, with $k \in \{0,\ldots,q-1\}$. Then, the second rectified cosine is simultaneously positive if $k$ is such that
\begin{align}
    -\frac{\pi}{2} \leq \frac{\pi}{2q} (k+m+\varepsilon) < \frac{\pi}{2},
\end{align}
where the upper limit is open because $\cos(\pi/2)=0$. Multiplying through by $2q/\pi$ and using the fact that $k+m$ and $q$ are integers, this means we should have
\begin{align}
    - q \leq k + m \leq q - 1. 
\end{align}
Therefore, the first interval of summation is truncated from $\{0,\ldots,q-1\}$ to 
\begin{align}
    I_{1} = \{\max(0,-m-q),\ldots,\min(q-1,q-m-1)\} , 
\end{align}
which is an empty set if $m > q$. 

Now consider the second range in the summation, with $k \in \{3q,\ldots,4q-1\}$. Then, the second rectified cosine is simultaneously positive if 
\begin{align}
    \frac{3\pi}{2} \leq  \frac{\pi}{2q} (k+m+\varepsilon) < \frac{5\pi}{2}; 
\end{align}
again the upper limit is open because $\cos(5\pi/2)=0$. This simplifies to
\begin{align}
    3q \leq k+m \leq 5q-1 , 
\end{align}
meaning that the second interval of summation is truncated from $\{3q,\ldots,4q-1\}$ to
\begin{align}
    I_{2} = \{\max(3q,3q-m), \ldots, \min(4q-1,5q-m-1)\},
\end{align}
which is empty if $m < -q$. 

We therefore at last have the summation
\begin{align}
    RSM_{s_1,s_{\alpha}} 
    &= \sum_{k\in I_{1} \cup I_{2}} \cos\left(\frac{\pi (k+\delta)}{2q} \right) \cos\left(\frac{\pi (k+m+\epsilon)}{2q} \right) ,  
\end{align}
which can be evaluated and simplified using Mathematica, yielding
\begin{align}
    RSM_{s_1,s_{\alpha}} &= \frac{1}{4} \sign(m) \csc\left(\frac{\pi}{2q}\right) \left[ \sin\left(\frac{\pi (m+1-\delta-\varepsilon)}{2q}\right) + \sin\left(\frac{\pi (m-1+\delta+\varepsilon)}{2q} \right) \right] \nonumber\\&\quad + \frac{2q-|m|}{2} \cos\left(\frac{\pi(m-\delta+\varepsilon)}{2q} \right) , 
\end{align}
using the convention $\sign(0) = 1$. This result is slightly unwieldy, but it is easy straightforward to verify that it satisfies several important sanity checks: it has the right large-$q$ limit, it agrees with our earlier result in the case $\alpha = \pi/2$ and $\alpha = \pi$, and it agrees with our result with $n=4$ and general $\alpha$.

With these checks out of the way, we now turn to the problem that the variables $m$ and $\varepsilon$, while convenient for computation, are not very useful if we want to apply the end result. To do so, we need to substitute in 
\begin{align}
    m = \left\lfloor \frac{2q}{\pi} (\varphi - \alpha) \right\rfloor = \left\lfloor \delta - \frac{2q}{\pi} \alpha \right\rfloor
\end{align}
and
\begin{align}
    \varepsilon = \frac{2q}{\pi} (\varphi - \alpha) - m = \delta - \frac{2q}{\pi} \alpha - \left\lfloor \delta - \frac{2q}{\pi} \alpha \right\rfloor. 
\end{align}
This does not lead to substantial simplification of the formula, but allows it to be evaluated numerically and compared to experiments.

\subsection{Amplitude variability }

We can also add other sources of RSM variability to the toy model. For example, we can let the amplitudes of the receptive fields vary, as would result from weights of fluctuating norm. Suppose in general that
\begin{align}
    h_{i}(x) = a_{i} \ReLU(w_{i}^{\top} x)
\end{align}
for some non-negative amplitude $a_{i}$, such that
\begin{align}
    RSM_{s_1,s_{\alpha}} 
    &= \sum_{k=0}^{n-1} a_{k}^{2} \ReLU\left( \cos\left(\frac{2\pi k}{n} + \varphi\right) \right) \ReLU\left(\cos\left(\frac{2\pi k}{n} + \varphi - \alpha \right) \right) . 
\end{align}
When all amplitudes are identical, then the RSM is a $2\pi/n$-periodic function of the gauge angle $\varphi$. This is because with equal amplitudes shifting the gauge angle by an integer multiple of $2\pi/n$ corresponds to circularly shifting the neurons, which does not change the RSM. Now, this circular shift also changes the amplitudes, and therefore can change the RSM. 

This can be easily seen by examining $RSM_{s_1,s_2}$ with $n=4$ neurons, for which we have
\begin{align}
    RSM_{s_1,s_2} = |\sin(\varphi) \cos(\varphi) | \times 
    \begin{cases}
        a_{1}^2, & 0 \leq \varphi < \pi/2  \\ 
        a_{4}^2, & \pi/2 \leq \varphi < \pi \\ 
        a_{3}^2, & \pi \leq \varphi < 3\pi/2 \\ 
        a_{2}^2, & 3\pi/2 \leq \varphi < 2\pi. \\ 
    \end{cases}
\end{align}

In general, we can proceed by writing
\begin{align}
    \varphi = \frac{2\pi}{n} (m + \delta)
\end{align}
in terms of $m \in \{0,\ldots,n-1\}$ and $\delta \in [0,1)$, and interpreting the indices of the amplitudes modulo $n$. Then, further assuming that $n=4q$, we can use the same argument before to obtain 
\begin{align}
    RSM_{s_1,s_2} 
    &= \sum_{k=0}^{q-1} a_{k-m}^{2} \cos\left(\frac{\pi (k+\delta)}{2q}\right) \sin\left(\frac{\pi(k+\delta)}{2q}\right)
\end{align}
and 
\begin{align}
    RSM_{s_{1},s_{1}} 
    &= \sum_{k=0}^{q-1} a_{k-m}^{2} \cos\left(\frac{\pi (k+\delta)}{2q} \right)^2 + \sum_{k=3q}^{4q-1} a_{k-m}^{2} \cos\left(\frac{\pi (k+\delta)}{2q} \right)^2 . 
\end{align}
In general, these expressions do not have simple closed forms. However, we can see that $RSM_{s_1,s_2}$ is now not an even function of $\delta$ with its maximum at $\delta = 1/2$, as was true with equal amplitudes.

\section{Reflection-symmetric tilings of the sphere in higher dimensions}\label{app:sphere}

Performing a similarly-detailed analysis of the RSMs for tiling representations of higher-dimensional manifolds is challenging. Even for the ordinary sphere $\mathbb{S}^{2}$, explicitly writing down a tiling solution for an arbitrary number of neurons is challenging---this corresponds to a variant of the classic Thompson problem in potential theory, to which the general solution remains unknown \citep{delbono2024most}. 

Though we cannot write down the RSM explicitly, we can characterize its overall structure for solutions with an even number of neurons tiling the sphere $\mathbb{S}^{d-1}$ in $d$ dimensions that obey a reflection-symmetry condition on the weights: for each neuron, there is another neuron with its receptive field oriented exactly opposite to the first. Let the number of neurons be $n=2N$, and assume that the stimulus-by-neuron weight matrix $W \in \mathbb{R}^{d \times 2 N}$ has the form
\begin{align}
    W = (U^{\top}, -U^{\top})
\end{align}
for some matrix $U \in \mathbb{R}^{N \times d}$.\footnote{Our results extend to larger networks where neurons are duplicated so long as the weights are appropriately normalized. We leave to future work a full investigation of when this form of the RSM applies to larger networks that are equivalent to these small networks thanks to further internal symmetries of the architecture \citep{simsek2021geometry,fukumizu2019embedding}.} Generalizing our study of the $d=2$ case, we take our probe stimuli to be the standard basis vectors and their negations, \textit{i.e.}, we have a stimulus matrix\footnote{As long as we assume that the test stimuli are given by a set of $d$ orthonormal vectors along with their negations, we lose no generality in making this choice because the global rotation symmetry implies that we can rotate the basis so that the test stimuli are axis-aligned.}
\begin{align}
    S = (I_{d}, -I_{d}) \in \mathbb{R}^{d \times 2d}, 
\end{align}
whose representation is 
\begin{align}
    H = \ReLU(W^{\top} S) = \begin{pmatrix} \ReLU(U) & \ReLU(-U) \\ \ReLU(-U) & \ReLU(U) \end{pmatrix} . 
\end{align}
We show below that, assuming weight tying, demanding that the probe stimuli are faithfully reconstructed implies that the matrix $U$ must be semi-orthogonal, \textit{i.e.},
\begin{align}
    U^{\top} U = I_{d}. 
\end{align}
The tiled representations of $\mathbb{S}^{1}$ that we considered before are a special case of such encodings (see Appendix \ref{sec:many_neuron_reconstruction}). 

If the above conditions hold, then the RSM has the general form 
\begin{align} \label{eqn:general_sphere_rsm}
    RSM = \begin{pmatrix} I_{d} + R & R \\ R & I_{d}+R \end{pmatrix}
\end{align}
where $R$ is a $d \times d$ symmetric matrix with zeros along the diagonal. This matches the structure we observed for $d=2$, where we saw that
\begin{align}
    RSM = \begin{pmatrix} 1 & \rho & 0 & \rho \\ \rho & 1 & \rho & 0 \\ 0 & \rho & 1 & \rho \\ \rho & 0 & \rho & 1 \end{pmatrix}
\end{align}
for a scalar $\rho$. 

The main salient feature of this matrix for $d>2$ is the fact that its elements obey non-trivial equality relations (relative to the case of $\mathbb{S}^{1}$, where all non-zero off-diagonal elements of the RSM were equal). For instance, if $d=3$, this yields an RSM of the form
\begin{align} \label{eqnRSM3dim}
    RSM = H^{\top} H = 
    \begin{pmatrix}
        1 & \rho_{1} & \rho_{2} & 0 & \rho_{1} & \rho_{2} \\ 
        \rho_{1} & 1 & \rho_{3} & \rho_{1} & 0 & \rho_{3} \\ 
        \rho_{2} & \rho_{3} & 1 & \rho_{2} & \rho_{3} & 0 \\ 
        0 & \rho_{1} & \rho_{2} & 1 & \rho_{1} & \rho_{2} \\ 
        \rho_{1} & 0 & \rho_{3} & \rho_{1} & 1 & \rho_{3} \\ 
        \rho_{2} & \rho_{3} & 0 & \rho_{2} & \rho_{3} & 1 
    \end{pmatrix} .
\end{align}
where $\rho_1$, $\rho_2$, and $\rho_3$ are the three distinct non-zero elements of the matrix $R$. The three functions $\rho_{1}$, $\rho_{2}$, and $\rho_{3}$ are continuous piecewise functions of the three gauge angles that appear in $d=3$, which can be computed explicitly using Mathematica---though their particular form is not illuminating. We see in Figure \ref{fig_supp_3_6_3} that the equality relations between different elements of the RSM implied by \eqref{eqnRSM3dim} are in fact obeyed to high accuracy in experiment. In Figure \ref{fig_10_20_10} we show that we observe the corresponding generalized structure \eqref{eqn:general_sphere_rsm} empirically in SGD-trained networks for $d=10$.

\subsection{Structure of the reconstruction}\label{app:sphere_reconstruction}

We first show that the matrix $U$ must be semi-orthogonal in order to faithfully reconstruct the probe stimuli $S$. If we assume weight-tying, we then have the reconstruction
\begin{align}
    \hat{S} = W \ReLU(W^{\top} S) = (A,-A)
\end{align}
where
\begin{align}
    A &= U^{\top} \ReLU(U) - U^{\top} \ReLU(-U) = U^{\top} \big[ \ReLU(U) - \ReLU(-U) \big]. 
\end{align}
But, as $\ReLU(x) - \ReLU(-x) = x$ for any $x \in \mathbb{R}$, this simplifies to
\begin{align}
    A = U^{\top} U. 
\end{align}
To exactly reconstruct the test stimuli, we should have $\hat{S}=S$ and thus $A = I_{d}$, which implies that the matrix $U$ must be semi-orthogonal:
\begin{align}
    U^{\top} U = I_{d}.
\end{align}
Moreover, perfect reconstruction is possible for \textit{any} semi-orthogonal $U$; choosing one such $U$ corresponds to choosing a gauge. 

\subsection{Structure of the RSM}

We now show that the RSM has the form \eqref{eqn:general_sphere_rsm}. For this representation, we have the RSM
\begin{align}
    RSM = H^{\top} H = \begin{pmatrix} Q & R \\ R & Q \end{pmatrix}
\end{align}
where we have defined the $d \times d$ blocks
\begin{align}
    Q &= \ReLU(U)^{\top}\ReLU(U) + \ReLU(-U)^{\top}\ReLU(-U)
    \\
    R &= \ReLU(U)^{\top} \ReLU(-U) + \ReLU(-U)^{\top}\ReLU(U),
\end{align}
Noting that $R$ is symmetric, to prove that the RSM takes the form \eqref{eqn:general_sphere_rsm} it suffices to show that $R$ has zeros along the diagonal and that $Q = U^{\top} U + R$. 

The fact that $R$ has zeros along the diagonal is easy to see upon expanding in indices: 
\begin{align}
    R_{ii} = \sum_{k=1}^{N} [\ReLU(U_{ki}) \ReLU(-U_{ki}) + \ReLU(-U_{ki}) \ReLU(U_{ki}) ] = 0, 
\end{align}
as $\ReLU(x)\ReLU(-x) = 0$ for all $x \in \mathbb{R}$. 

To show that $Q = U^{\top} U + R$, we observe that for $x,y \in \mathbb{R}$ we have
\begin{align}
\begin{split}
    \ReLU(x) \ReLU(y) + \ReLU(-x) \ReLU(-y) - \ReLU(x) \ReLU(-y) - \ReLU(-x) \ReLU(y) 
    = xy . 
\end{split}
\end{align}
Applying this identity element-wise, we have
\begin{align}
    Q - R &= \ReLU(U)^{\top}\ReLU(U) + \ReLU(-U)^{\top}\ReLU(-U) \nonumber\\&\quad -  \ReLU(U)^{\top} \ReLU(-U) - \ReLU(-U)^{\top}\ReLU(U)
    \\
    &= U^{\top} U. 
\end{align}
This proves that $Q = U^{\top} U+R$, and thus that in general we have
\begin{align}
    RSM = \begin{pmatrix} U^{\top} U+ R & R \\ R & U^{\top} U+ R \end{pmatrix}. 
\end{align}
Note that this applies for any set of weights obeying the reflection-symmetry condition. For a solution satisfying $U^{\top} U = I_{d}$, this reduces to \eqref{eqn:general_sphere_rsm}. As $R$ is a $d \times d$ symmetric matrix with zeros along the diagonal, it has $d(d-1)/2$ independent elements, matching the dimensionality of the gauge group.

\section{\texorpdfstring{\citet{cui2022deconfounded}}{Cui et al.}'s deconfounding method does not eliminate gauge dependence} \label{app:cui}

\citet{cui2022deconfounded} have proposed a method to deconfound CKA by linearly regressing out components of the RSM that are inherited from the input-space kernel. This can help correct for the possibility that two representations appear spuriously similar because they share an input-driven component. Here, we show that this method cannot in general remove the gauge-dependence of the RSM, even if one can measure the population kernel (\textit{i.e.}, all possible stimulus pairs). For any pair of stimuli $x,x'$ and any group element $g$, let $K_{g}(x,x') = K(g^{-1}\cdot x, g^{-1} \cdot x')$ be the representation-space kernel after a gauge transformation. Let $K^{0}$ be the input-space kernel, which we assume respects the group structure in the sense that $K^{0}(g\cdot x,g \cdot x') = K^{0}(x,x')$. \citet{cui2022deconfounded} propose to consider 
\begin{align}
    dK_{g}^{1} = K_{g}^{1} - \alpha_{g} K^{0}, \quad \textrm{where} \quad \alpha = \frac{\langle K^{0}, K_{g}^{1} \rangle}{\Vert K^{0} \Vert^2},
\end{align}
in terms of the Frobenius inner product on kernel matrices. At the population level, 
\begin{align}
    \langle K^{0},K^{1} \rangle = \mathbb{E}_{x,x'}[K^{0}(x,x') K_{g}^{1}(x,x')],
\end{align}
and similarly $\Vert K^{0} \Vert^2 = \mathbb{E}_{x,x'}[K^{0}(x,x')^2]$. Now, as the data distribution and input kernel are $G$-invariant, using the definition of $K_{g}^{1}$ we see that
\begin{align}
    \alpha = \frac{\langle K^{0}, K_{g}^{1} \rangle}{\Vert K^{0} \Vert^2} = \frac{\langle K^{0}, K^{1} \rangle}{\Vert K^{0} \Vert^2},
\end{align}
\textit{i.e.}, that $\alpha$ is gauge-invariant. As a result, before projection onto the PSD cone, for any two gauges $g$ and $g'$ we will have 
\begin{align}
    dK_{g}-dK_{g'} = K_{g}-K_{g'};
\end{align}
the proposal of \citet{cui2022deconfounded} subtracts off a gauge-invariant constant matrix. Therefore, it cannot in general eliminate the gauge-dependence of the RSM.

\section{Additional experimental details}\label{app:experiments}

\subsection{Toy setup}\label{sec:appendix:toysetup}

Our toy setup consists of localized RFs on a circle that are created by responses of $n$ ReLU neurons.
Specifically, the response of each neuron is $h_i(x) = \text{ReLU}(w_i^{\top}x - b_i)$, where $w_i = |w_i|(\cos(\theta_i), \sin(\theta_i))$ is the incoming weight defined by $\theta_i$, and $b_i$ is the bias for that neuron. To allow for RFs to have different amplitudes ($a_i$) and tuning width ($\delta_i$), we set these values accordingly. 
Given that stimuli lie on a circle, $x = (\cos(\theta), \sin(\theta))$, we have: $h_i(x) = \text{ReLU}(|w_i|\cos(\theta-\theta_i) - b_i)$. The conditions for amplitude and tuning width are $|w_i| - b_i = a_i$ and $|w_i|\cos(\delta_i/2)-b_i=0$, respectively. Solving for $|w_i|$ and $b_i$ leads to:
\begin{align}
|w_i| = \frac{a_i}{1-\cos(\frac{\delta_i}{2})}, \quad b_i = \frac{a_i\cos(\frac{\delta_i}{2})}{1-\cos(\frac{\delta_i}{2})} 
\end{align}
Note that in \S\ref{sec:toy} we had $a_i=1$ and $\delta_i = \pi$, which leads to $b_i=0$ and $|w_i|=1$. However, in \S\ref{sec:factors} we used arbitrary amplitudes and tuning widths.

\paragraph{Approximate rescaling of RFs under different tuning widths}

We remark that in this setting changing the receptive field widths (and number of neurons, to maintain uniform coverage) is nearly, but not exactly, equivalent to re-scaling of the input space. This approximate equivalence follows from the fact that the receptive field shapes change slightly depending on their width. Concretely, consider a receptive field centered at $\theta = 0$, of width $\delta$. Its response as a function of angle is
\begin{align}
    h(\theta,\delta) = \ReLU\left(1 - \frac{1-\cos(\theta)}{1-\cos(\delta/2)} \right). 
\end{align}
For changing the width to be equivalent to a re-scaling of the input, for any new narrower width $\delta'< \delta$ we would like to find a scale factor $\alpha$ such that
\begin{align}
    h(\theta,\delta) \approx h(\alpha \theta, \delta') . 
\end{align}
This is clearly not feasible globally, as the function
\begin{align}
    \frac{1-\cos(\theta)}{1-\cos(\delta/2)} = \left(\frac{\sin(\theta/2)}{\sin(\delta/4)}\right)^2
\end{align}
is not positive-homogeneous in $\theta$. However, we can approximately achieve this locally. Near $\theta = 0$, the desired approximation holds to second order in $\theta$ if we take
\begin{align}
    \alpha = \sqrt{\frac{1-\cos(\delta'/2)}{1-\cos(\delta/2)}}. 
\end{align}
This incurs an error near the edges of the receptive field, where the re-scaled receptive field has width
\begin{align}
    \frac{\sin(\delta/4)}{\sin(\delta'/4)} \delta'
\end{align}
This is an increasing function of $\delta'$, and is equal to $\delta$ at $\delta' = \delta$. Therefore, in general with this local approximation there will be some discrepancy between the re-scaled receptive field and the original one. However, this discrepancy is small; if $\delta = \pi$ then the relative width attains a minimal value of $\simeq 0.9003$ as $\delta' \to 0$. 

Another way to see the near-equivalence of changing the receptive field width and re-scaling the input is by matching the widths of the RFs. Then, one should choose $\alpha = {\delta'}/{\delta}$, which produces a small discrepancy for intermediate values of $\theta$. This can be seen by observing that the ratio of the coefficients of the quadratic terms of the Taylor expansions of $h(\theta,\delta)$ and $h(\delta'\theta/\delta,\delta')$ is
\begin{align}
    \left( \frac{\delta' \sin(\delta/4)}{\delta \sin(\delta'/4)} \right)^2 . 
\end{align}
This ratio is nothing but the square of the ratio of widths we found before; it is again an increasing function of $\delta'$ that is close to unity. For instance, with $\delta = \pi$ it is minimized by taking $\delta' \to 0$, where it takes value $\simeq 0.8106$. 

Together, these observations give intuition for the approximate collapse of curves we saw in Figure \ref{fig2}: the RSM variability for different tuning widths and different numbers of neurons should approximately depend only on the product of the tuning width and the number of neurons because changing the width is \emph{nearly} equivalent to expanding the input space. 

\subsection{Neural network simulations}
Here we provide experimental details regarding the simulations provided throughout the paper. Neural networks consisted of a two-layer autoencoder with input dimension $d$, hidden-layer with ReLU activation consisting of $n$ neurons, and an output layer that was a reconstruction of the input ($y=x$). Networks were trained using a vanilla SGD with a fixed learning rate ($\eta$), a weight decay ($\gamma$) and varying batch size ($b$). In some simulations, an $L_1$ penalty on the activations of the hidden layer was imposed. Training was performed on PyTorch with NVIDIA GeForce RTX 2080 Ti GPU.  Specifics parameters are mentioned for each simulation below. 
\subsubsection{Simulations with Gaussian data}
Data were drawn in an online way from normal $d$-dimensional standard Gaussian distribution. Prior to the online training, a warm-up pretraining was performed with no weight decay and batch size of 128 (this corresponds to time 0 on the axes of the training plots). The plots of RSM values over time were created by first saving snapshots of the model at intervals of 100 time steps, and then applying a moving average filter with the window size of 5. The specific parameters for each figure are as follows:

Figure \ref{fig_sgd_242}: $d=2$, $n=4$, $\eta=0.1$, $\gamma=0.1$, $b=1$.

Figure \ref{fig4}: $d=2$, $n=15$, $\eta=0.15$, $\gamma=0.1$, $b=1$.

Figure \ref{fig_supp_init_2_15_2_l1}:
$d=2$, $n=15$, $\eta=0.1$, $\gamma=0.1$, $b=100$, $\lambda_{1} = 0.001$ (L1-penalty coefficient).

Figure \ref{fig_supp_3_6_3}: $d=3$, $n=6$, $\eta=0.1$, $\gamma=0.1$, $b=1$.

Figure \ref{fig_supp_init_2_4_2}: $d=2$, $n=4$, $\eta=0.1$, $\gamma=0.1$, $b=1$.

Figure \ref{fig_supp_init_2_15_2}: $d=2$, $n=15$, $\eta=0.1$, $\gamma=0.1$, $b=100$.

Figure \ref{fig_10_20_10}: $d=10$, $n=20$, $\eta=0.075$, $\gamma=0.05$, $b=1$.

\subsubsection{Simulations with rotated image data (Figures~\ref{fig_real} and \ref{fig_realdata_supp})}\label{app:real_methods}

As mentioned in \S\ref{sec:real} of the main text, here the data consists of rotated versions of a digit from the Kuzushiji-MNIST dataset \citep{clanuwat2018deep}. Specifically, we take one digit with dimensions $28 \times 28$ and create 180 rotated versions of it covering the full circle. The baseline image was a digit ($index=1$) from Kuzushiji-49 training dataset, and was accessed from \url{https://github.com/rois-codh/kmnist}, where it is available under a CC BY-SA 4.0 License.\footnote{Compared to MNIST digits, this image contained fewer intrinsic symmetries, which led to a smaller degree of interference with the latent symmetry. For example, using the numeral ``0'' would include a four-fold intrinsic symmetry of the digit.} 

The image data were flattened and fed into a two-layer autoencoder with ReLU nonlinearity and trainable bias. The input/output and the hidden layer dimensions were 784 and 32 respectively. Each model was trained using SGD with batch size $b=24$, learning rate $\eta=0.01$, weight decay $\gamma=0.005$, and $L_1$ penalty coefficient $2 \times 10^{-5}$ on hidden layer activations. Additionally, a small Gaussian synaptic noise was added to the weights for regularization (noise variance $\eta b \sigma^2$ for $\sigma=0.02$). The training was continued for $500$ epochs which is long after the training loss stabilized. 

Overall, these parameters allowed us to create approximately localized RFs, although this is not the only way to obtain such receptive fields. Finally, 1000 runs of the above training was performed with random initializations to achieve multiple instances of RFs. These simulations led to different number of RFs, as shown in the histogram in Fig.~\ref{fig_realdata_supp}. Nevertheless, the findings were consistent across all numbers. The results in Fig.~\ref{fig_real} of the main text correspond to models with $n=15$ RFs. 
Finally, to infer $\varphi$, a max projection of active RFs was performed to find the RF envelopes, and the phase of the dominant frequency of the envelope was chosen as the gauge variable for that run. This is illustrated in Fig.~\ref{fig_realdata_method}.

\subsection{Pretrained vision models (Figure~\ref{fig_realdata_largemodel})}
We further studied representations of the rotated image data in three exemplary pretrained vision models.
As convolutional baselines, we used ResNet-18, a standard residual CNN with approximately 11.7M parameters \citep{he2016deep}, and ConvNeXt-Tiny, a more modern convolutional architecture with roughly 28.6M parameters \citep{liu2022convnet}. For both CNNs, we analyzed the final-stage, global-average-pooled feature representations. As a transformer-based model, we used DINOv2-S/14-reg, which is a self-supervised Vision Transformer with 12 transformer blocks, 384-dimensional embeddings, $14 \times 14$ patches, and register tokens, with roughly 21M parameters; representations were taken from the final-block class token (\texttt{block\_11}) \citep{oquab2023dinov2}. The inputs are 360 rotated versions of the baseline image in Fig.~\ref{fig_real}a (from the Kuzushiji-MNIST dataset), color-coded by their rotation angle. ResNet-18 and ConvNeXt-Tiny pretrained weights were obtained through \texttt{torchvision.models} using the default weight configurations, and pretrained DINOv2-S/14-reg weights were obtained from the Torch Hub repository.

\clearpage 

\section{Additional figures}
\begin{figure}[h]
 \centering
\includegraphics[width=0.75\linewidth]{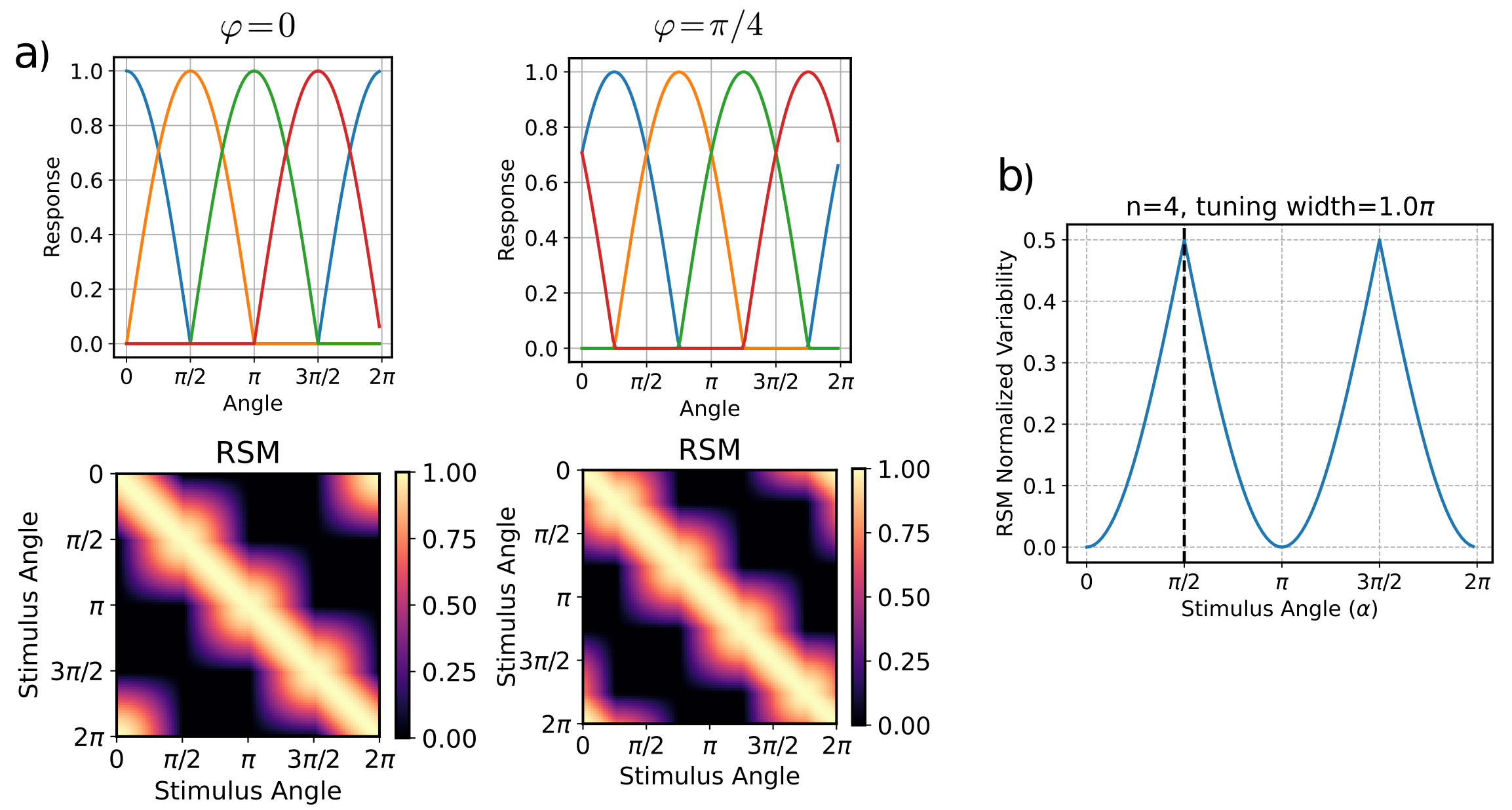}
\vspace{-1em}
    \caption{Stimulus-dependency of RSM and its dependence on $\varphi$. a) Examples of full-RSM matrices for two values of $\varphi=0$ and $\varphi=\pi/4$. b) Dependency of normalized RSM variability ($\Delta$) on the stimulus angle. Here, the RSM are calculated between $s_1=(1,0)^{\top}$ and $s_{\alpha} = (\cos{\alpha},\sin{\alpha})^{\top}$. All cases are for $n=4$ with tuning width of $\pi$, which is the same as the toy model in Fig.~\ref{fig1}.}
\label{fig_stim_supp}
\end{figure}

\begin{figure}[h]
 \centering
\includegraphics[width=0.5\linewidth]{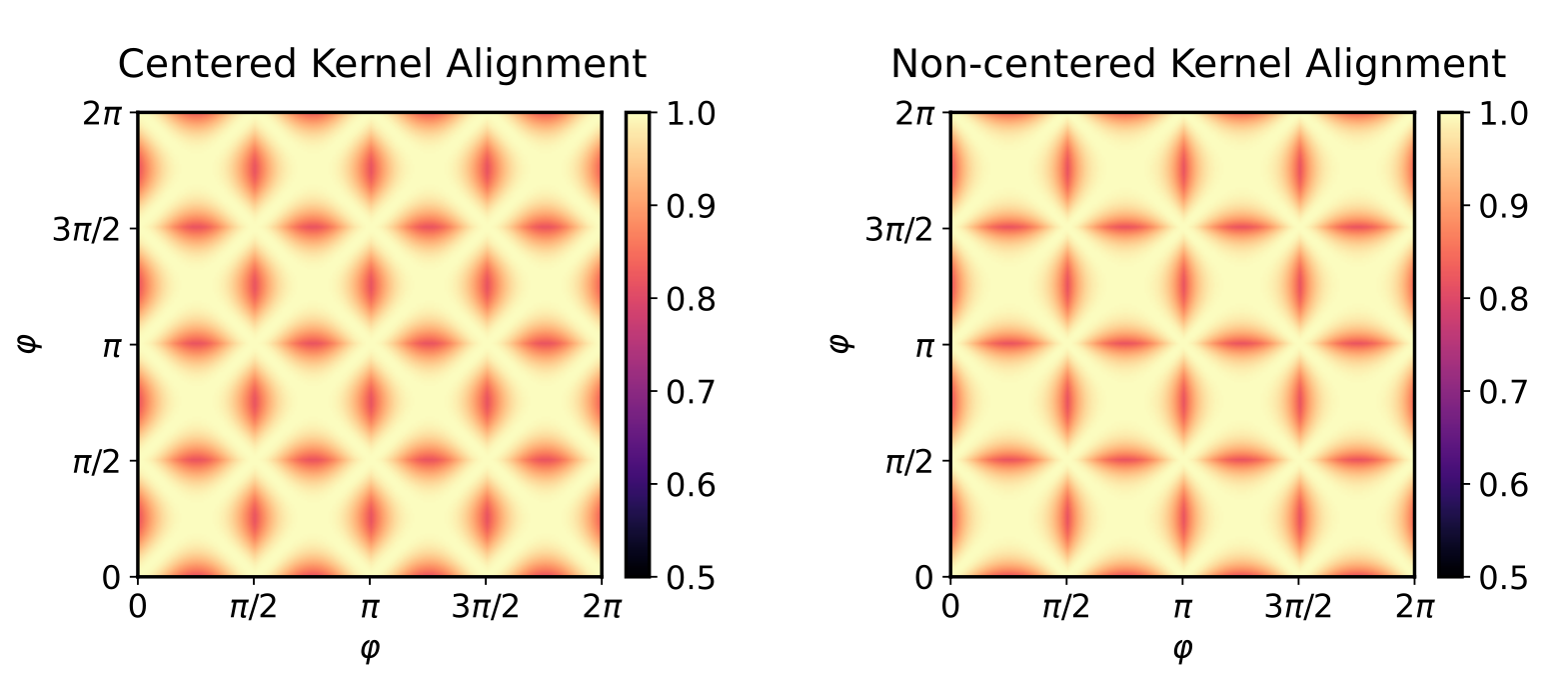}\vspace{-1em}
    \caption{Kernel alignment measures inherit gauge-dependency from RSM. 
    (Left) Centered Kernel Alignment (CKA) and (Right) non-centered kernel alignment between pairs of RSM matrices at different values of gauge variable. The values are calculated based on the toy setup of Figure \ref{fig1} with $n=4$ neurons and four trial stimuli. See also Appendix~\ref{app:4neuroncka}.}
\label{fig_cka_supp}
\end{figure}

\begin{figure}[h]
\hspace{-1em}
 \centering
\includegraphics[width=0.5\linewidth]{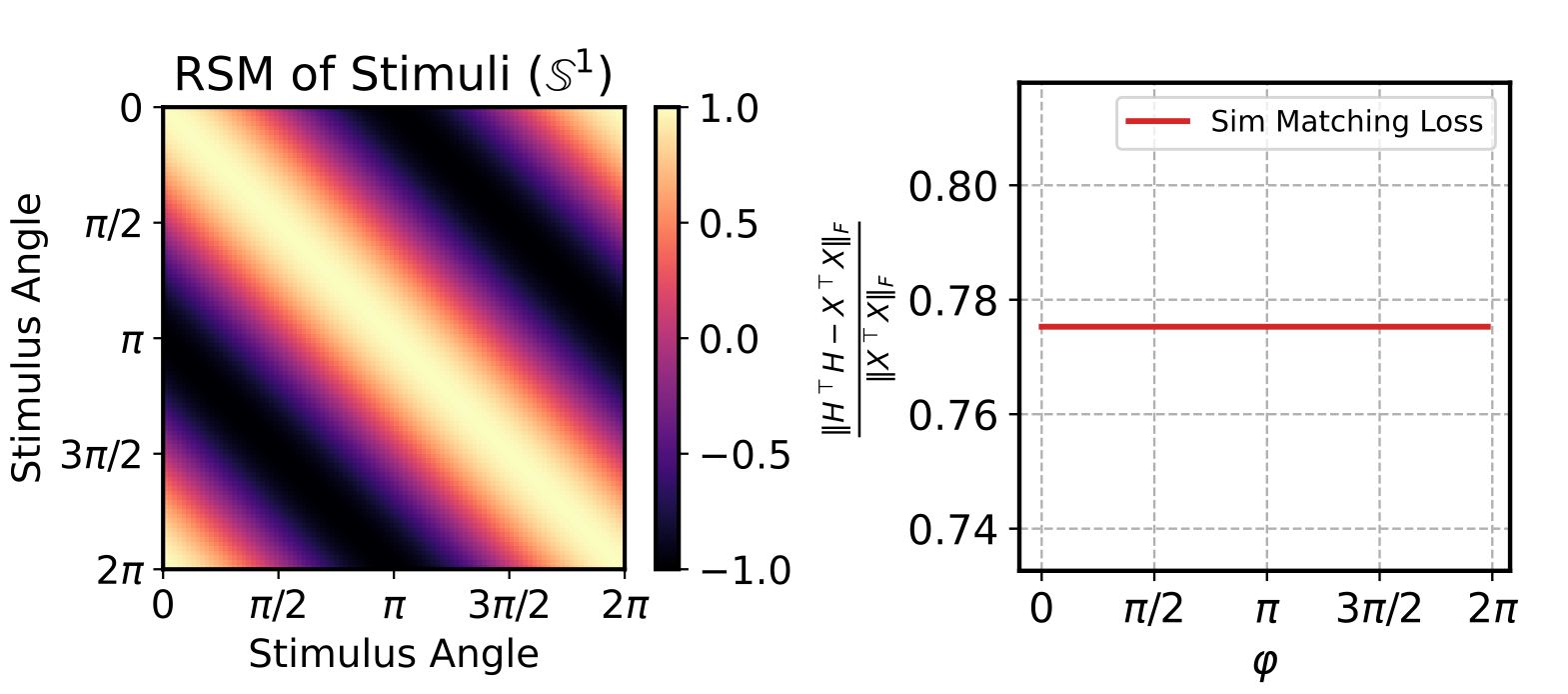}
\vspace{-1em}
    \caption{Gauge-dependence of RSM does not affect similarity matching loss.
    (left) Stimulus RSM $X^{\top}X$ on a ring $(\mathbb{S}^1)$. (right) Similarity matching loss $\frac{\|H^{\top}H - X^{\top}X\|_F}{\|X^{\top}X\|_F}$ as a function of gauge variable (setup is similar to Fig.~\ref{fig_stim_supp}, and RSM is calculated with 100 trial stimuli). We prove that this holds more generally in Appendix~\ref{app:simmatch}.}
\label{fig_simmatching_supp}

\end{figure}

\begin{figure}[h]
 \centering
\includegraphics[width=0.8\linewidth]{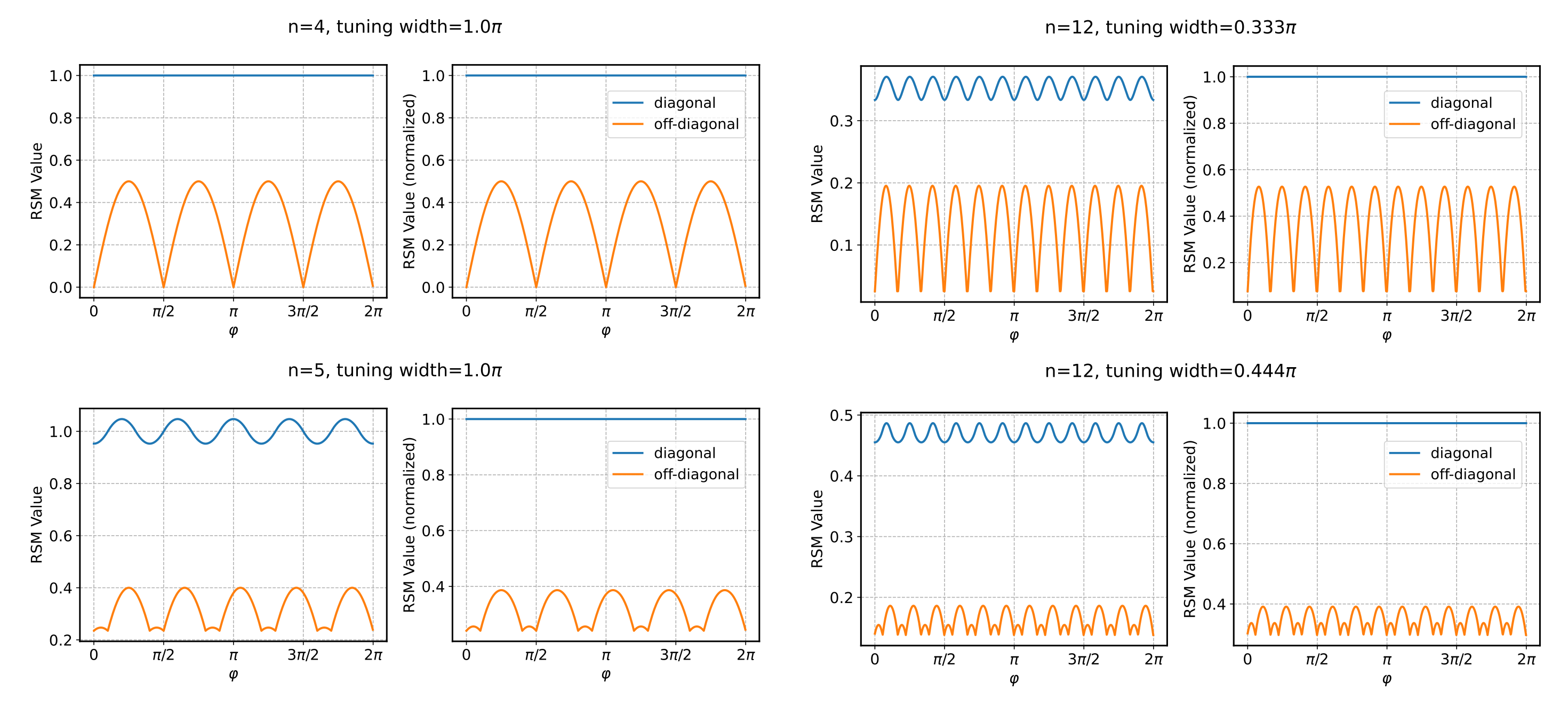}
\vspace{-1em}
    \caption{Additional examples of $\varphi$-dependency of RSMs for different numbers of RF and tuning widths. In all cases, the off-diagonal RSM is calculated for stimulus $s_1=(1,0)^{\top}$ and $s_{\alpha} = (\cos{\alpha},\sin{\alpha})^{\top}$, for $\alpha$ half of the tuning width.}
\label{fig_phicurves_supp}
\vspace{0em}
\end{figure}

\begin{figure}[h]
 \centering
\includegraphics[width=0.8\linewidth]{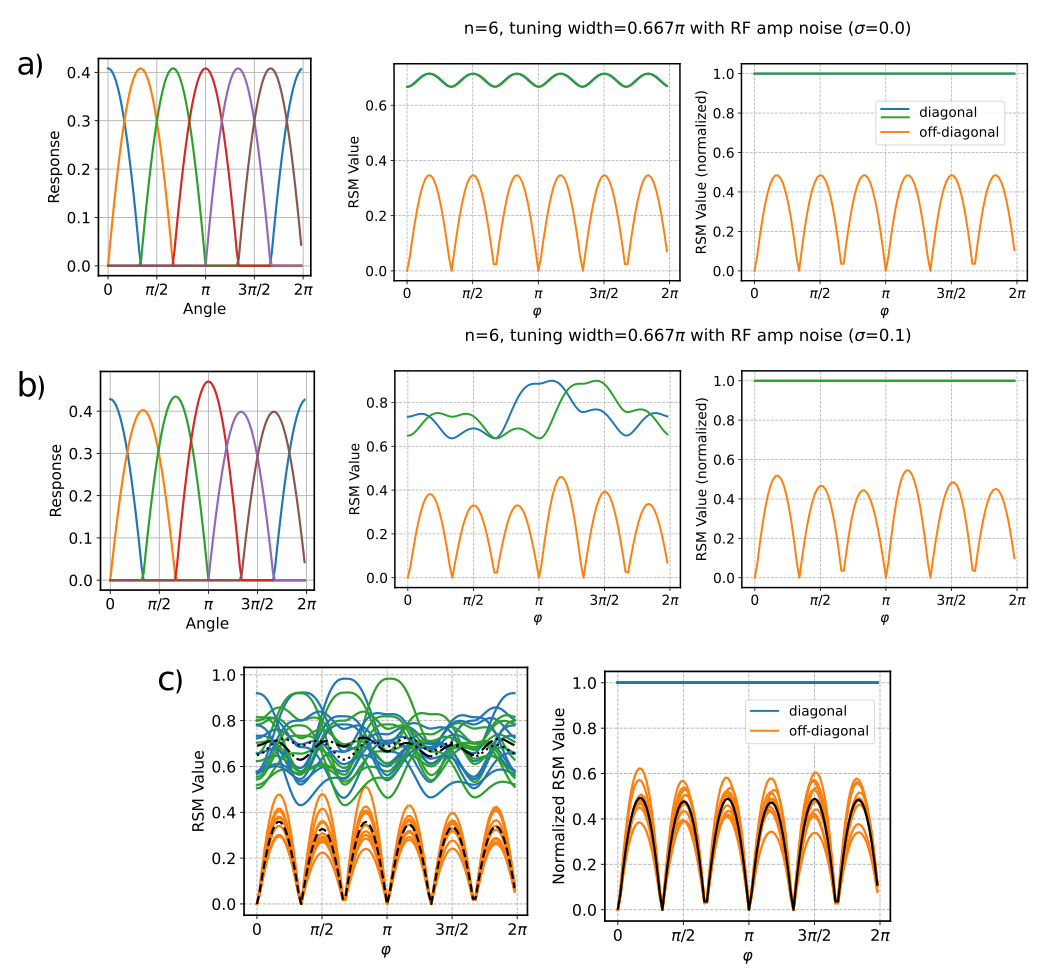}
    \caption{Additional results on RSM variability in neural codes with RF amplitude noise. Amplitude noise is added to existing RF as Gaussian noise with standard deviation $std = \sigma \times amp$, where $amp$ is the baseline amplitude. a) $\sigma=0$. b) $\sigma=0.1$. c) Average curves for 100 realizations of amplitude noise. In all cases, the blue and green curves correspond to diagonal elements corresponding to stimuli $s_1=(1,0)^{\top}$ and $s_{\alpha} = (\cos{\alpha},\sin{\alpha})^{\top}$ for $\alpha$ half of the tuning width, respectively. Similarly, the orange off-diagonal curves show RSM entries between $s_1$ and $s_{\alpha}$.}
\label{fig_amp_supp}
\vspace{0em}
\end{figure}

\begin{figure}[t!]
    \centering
    \includegraphics[width=0.65\linewidth]{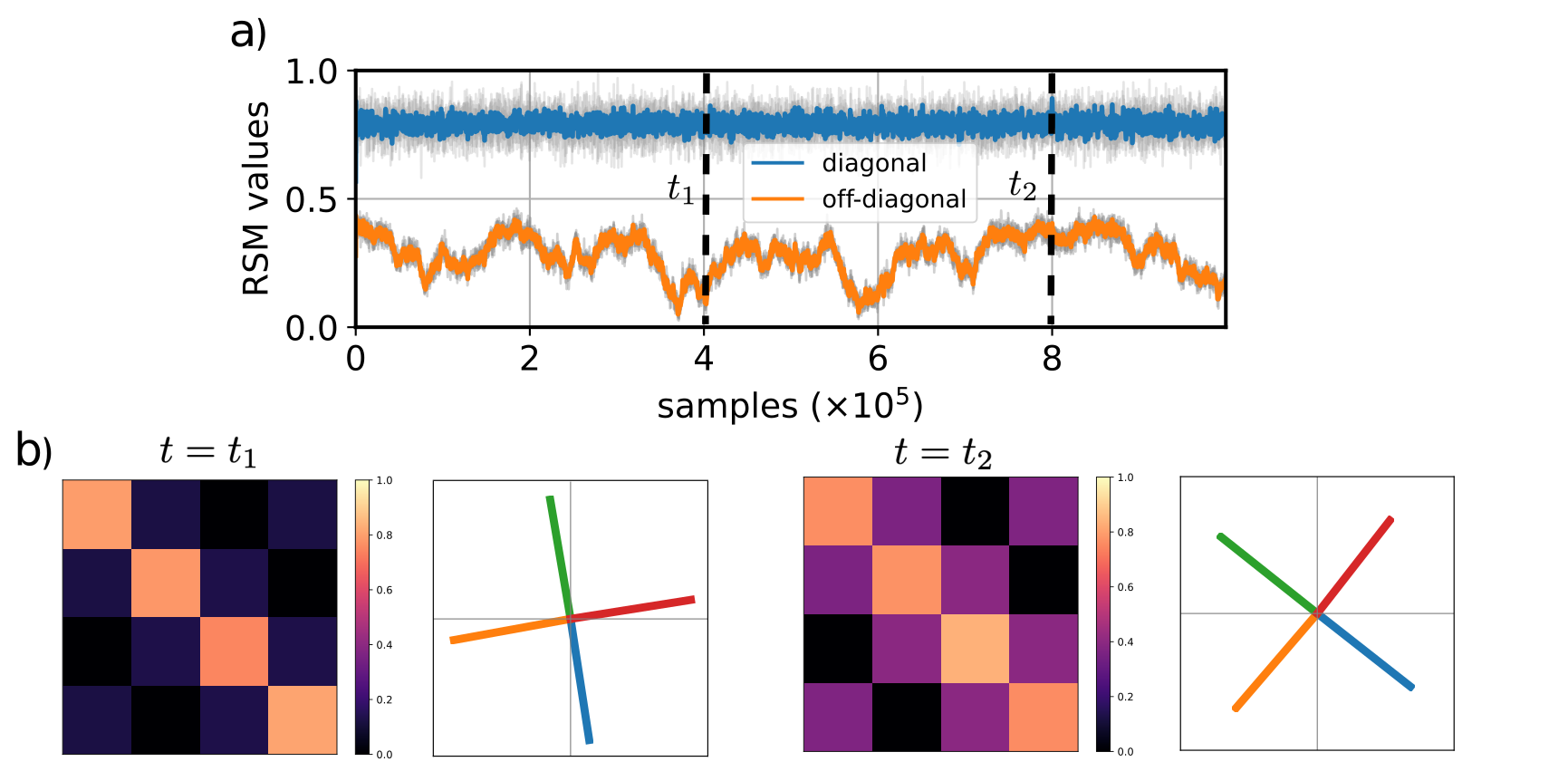}
    \caption{\text{Continual training under SGD leads to a variable RSM}. Input is two-dimensional and there are $n=4$ ReLU neurons. a) Values of RSM over time. b) At two time snapshots of training, the RSM matrices (left), and the corresponding weight vectors (right) are shown.}
    \label{fig_sgd_242}
\end{figure}

\begin{figure}[h]
    \centering
\includegraphics[width=0.6\linewidth]{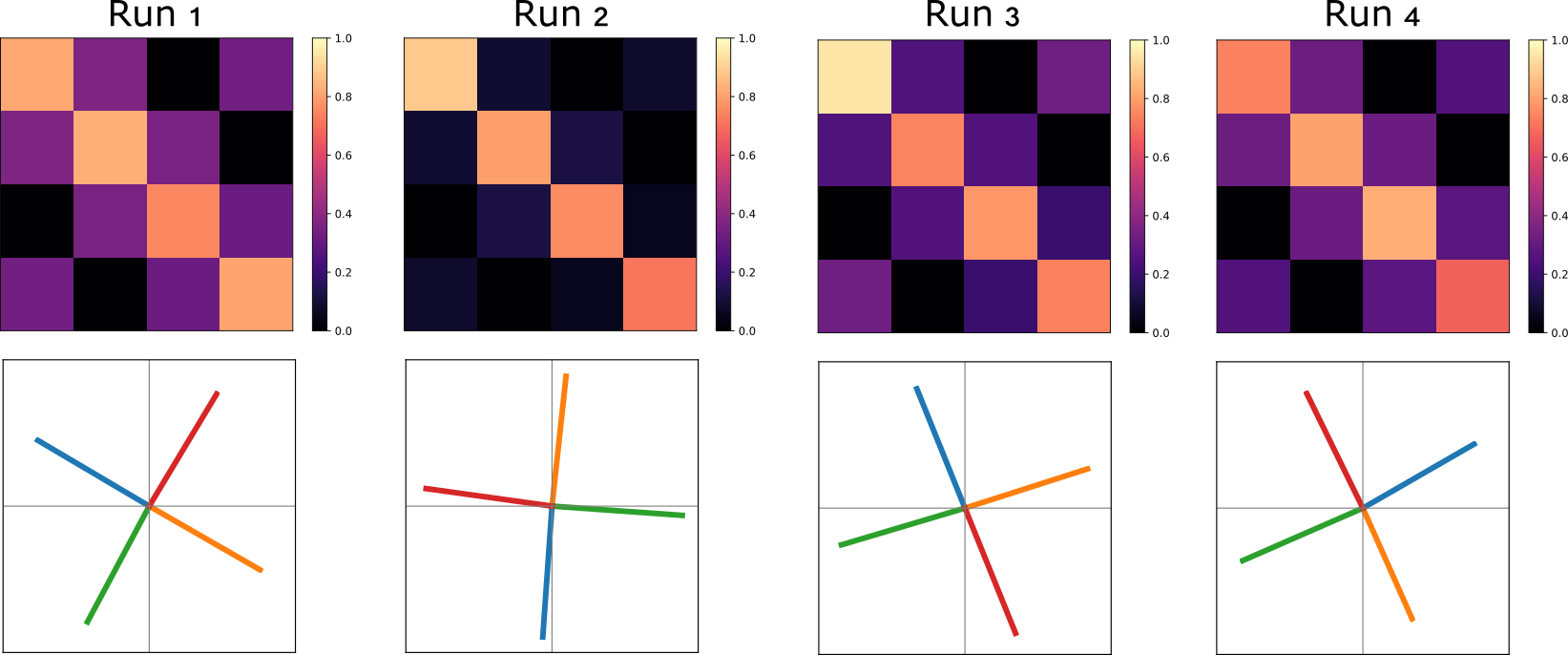}
    \caption{RSM plots (top) and neurons' weight vectors (bottom) for four different runs and $n=4$ neurons. Each simulation is run with $5 \times 10^4$ samples seen (batch size of 1).}
\label{fig_supp_init_2_4_2}
\vspace{-1em}
\end{figure}

\begin{figure}[h]
    \centering
\includegraphics[width=0.6\linewidth]{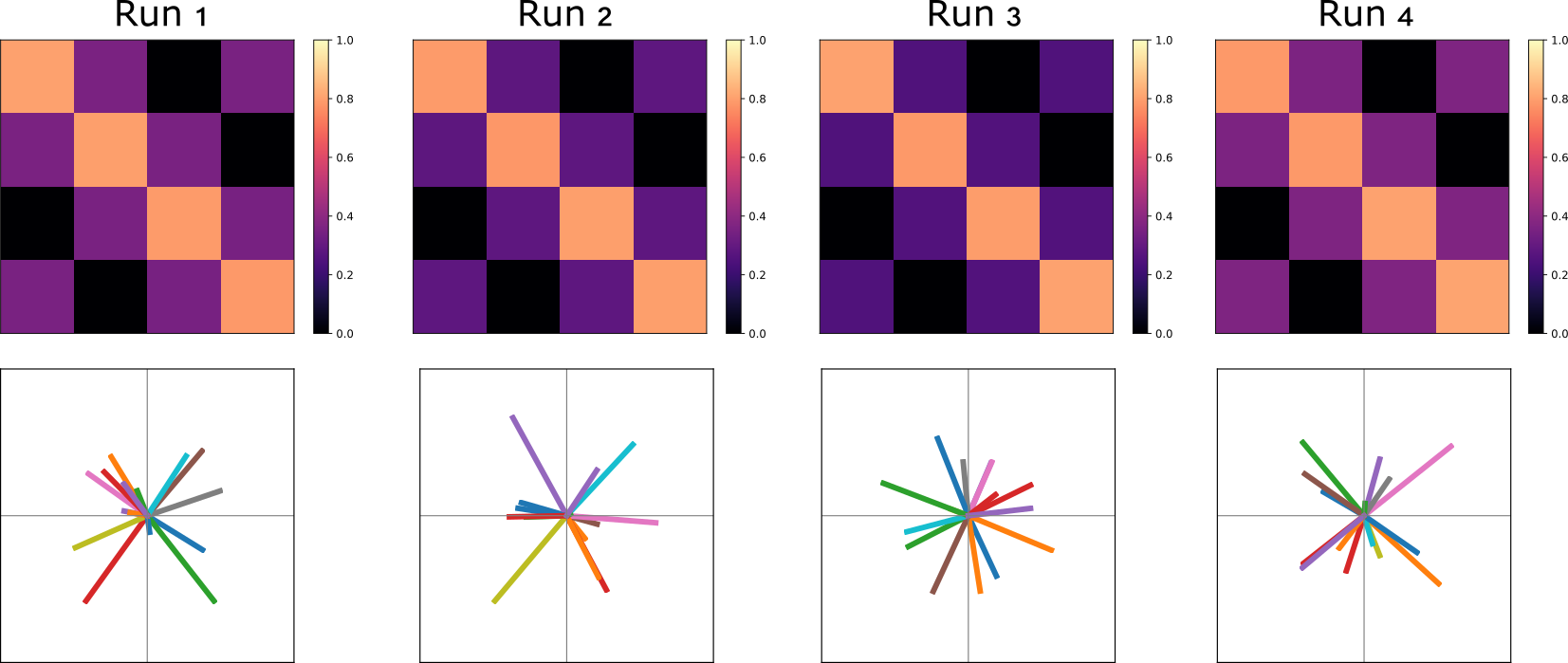}
    \caption{RSM plots (top) and neurons' weight vectors (bottom) for four different runs and $n=15$ neurons. The batch size is large, leading to a relatively low SGD-noise regime where, unlike Fig.~\ref{fig4} in the main text, neurons' weights do not collapse. Each simulation is run with $5 \times 10^4$ samples seen (batch size of 100).}
\label{fig_supp_init_2_15_2}
\vspace{0em}
\end{figure}

\begin{figure}[h]
    \centering
\includegraphics[width=0.9\linewidth]{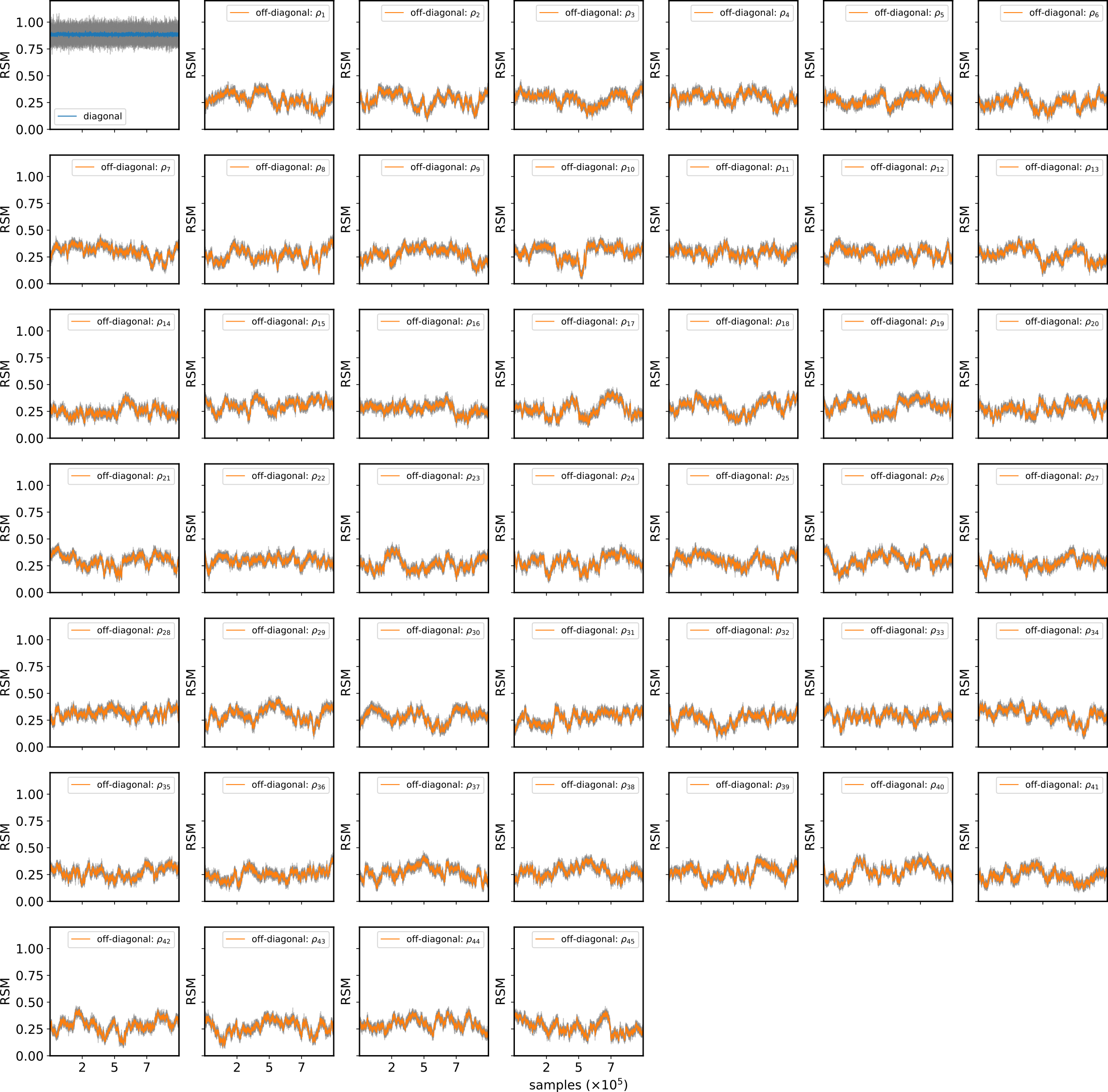} \caption{RSM values over time for a network with input dimension $d=10$ and $n=20$ neurons. Network is trained with SGD and batch size of one. Based on the predictions in Appendix \ref{app:sphere}, the non-zero off-diagonal elements are placed into $d(d-1)/2 =45$ groups. Each group contains 4 RSM values which, as predicted, are highly correlated (gray curves: members of the group, orange: group mean).}
\label{fig_10_20_10}
\vspace{3em}
\end{figure}

\begin{figure}[h]
 \centering
\includegraphics[width=0.55\linewidth]{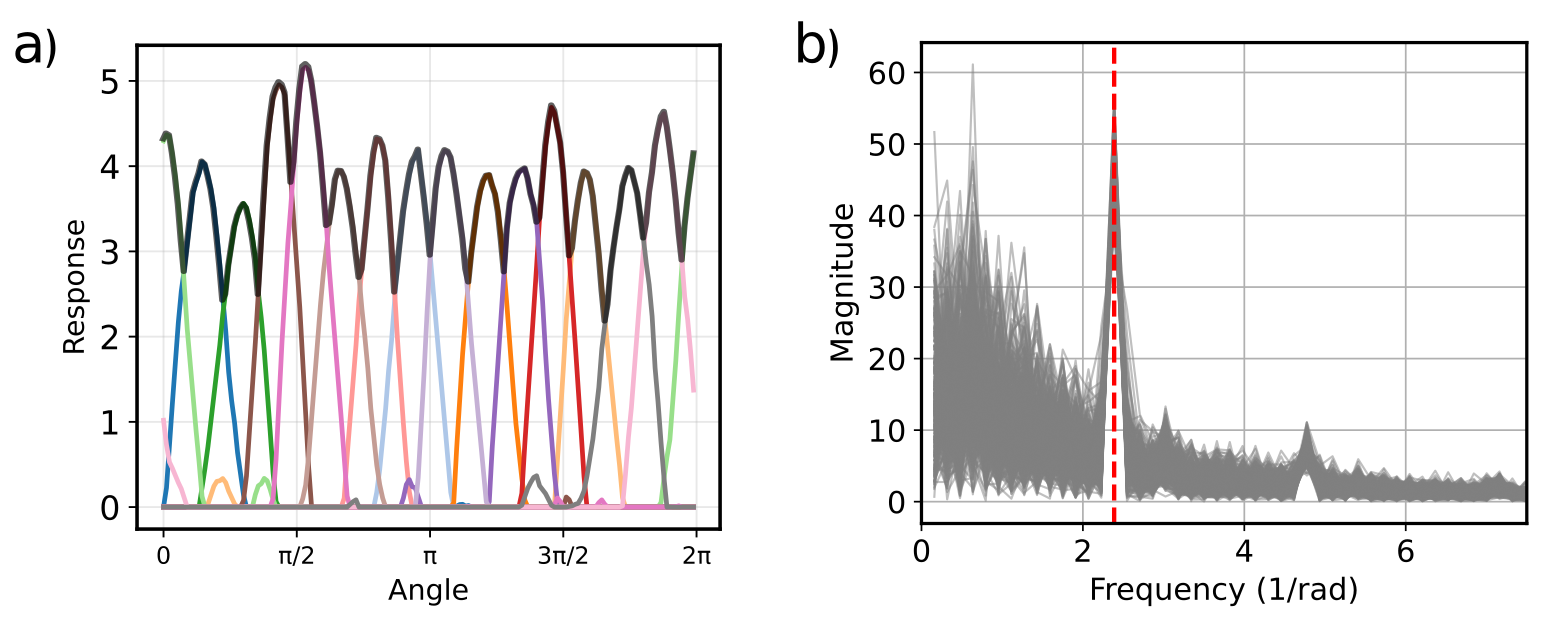}
    \caption{Inference of the gauge variable $\varphi$ from RF profiles of trained networks. a) An example of a run with $n=15$ RFs. Black line shows the envelope of the RFs found by max projection. b) Frequency spectrum of the RF envelopes for multiple runs. For each run, $\varphi$ is inferred by finding the phase at the dominant frequency at $n/2\pi$ (shown by red dashed line). }
\label{fig_realdata_method}
\vspace{0em}
\end{figure}

\begin{figure}[h]
 \centering
\includegraphics[width=0.8\linewidth]{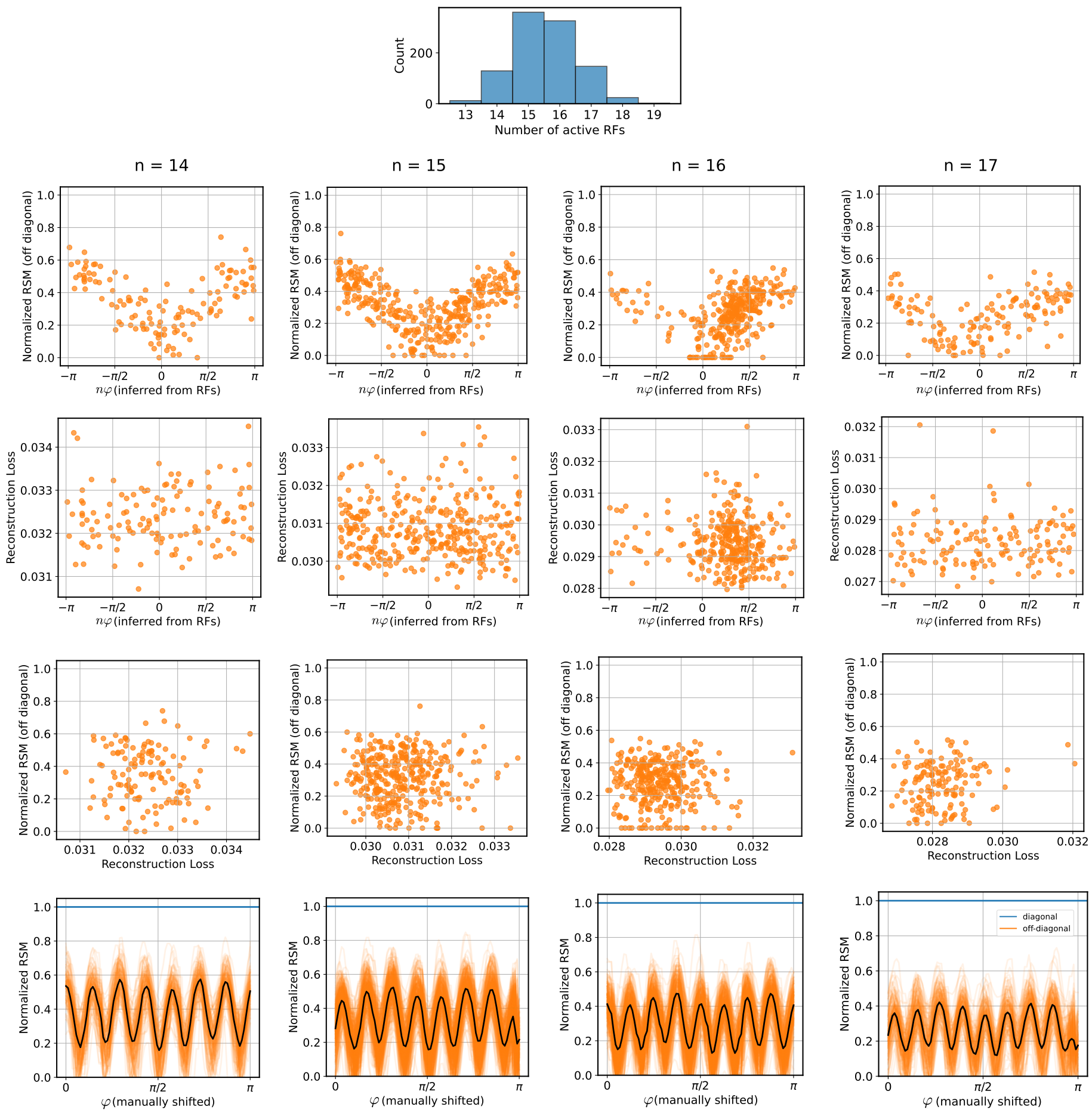}
    \caption{Additional results on RSM variability in real-world data (see Fig.~\ref{fig_real} and the main text for details of the plots). 1000 realizations of the autoencoder were trained on rotated versions of a given image. The histogram on top shows the distribution of the number of RFs across runs. For each number, the associated plots of RSM variability are shown in a column. The $\varphi$-dependency of RSMs is evident irrespective of the number of RFs. In all plots, the orange data correspond to the off-diagonal entry of the RSM and are calculated for two stimuli that are $\alpha=25^{\cdot}$ apart.}
\label{fig_realdata_supp}
\vspace{0em}
\end{figure}

\begin{figure}[h]
 \centering
\includegraphics[width=0.5\linewidth]{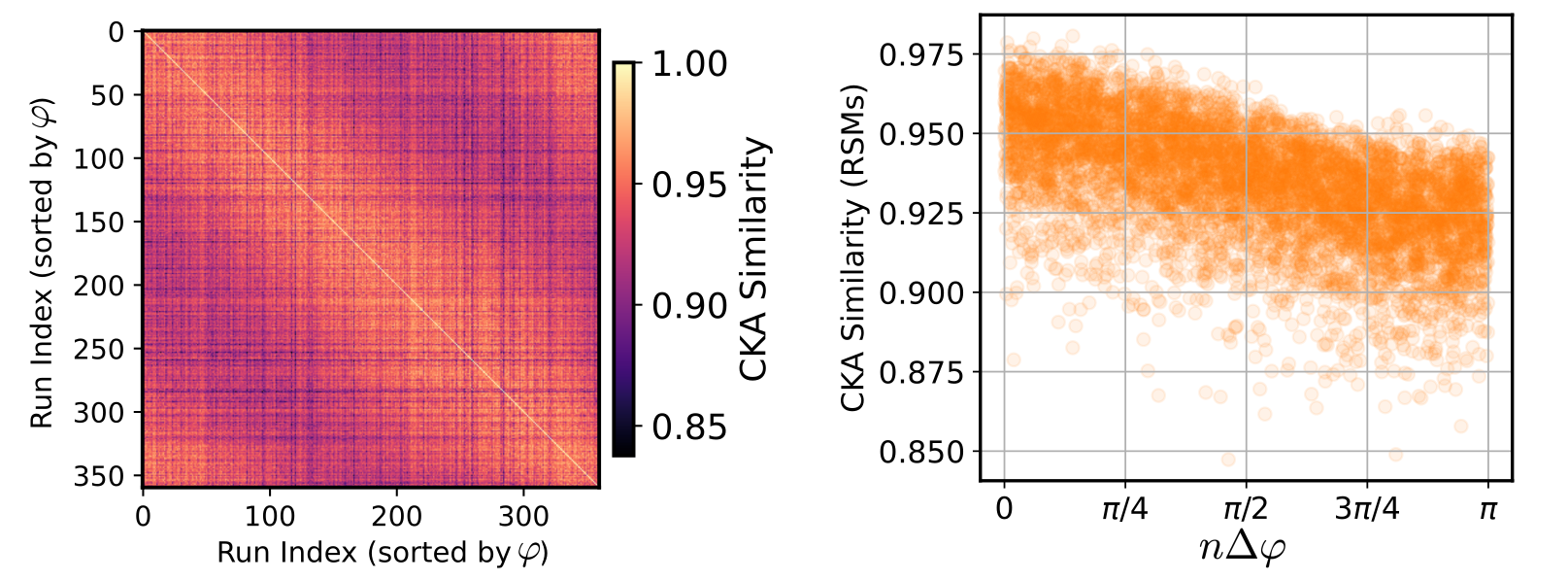}
    \caption{Additional results on CKA analysis on autoencoding image data with latent symmetry in Section \ref{sec:real}. The analysis is similar to panels g and h in Figure \ref{fig_real}, but CKA similarity is measured from non-normalized RSMs. (Left) CKA similarity matrix calculated between RSMs of different runs. The runs are sorted based on the inferred $\varphi$. (Right) Scatter of CKA similarity as a function of gauge difference (each point represents a pair of runs, and for visualization purposes only $1/10$ of the data are shown; Pearson's $r=-0.56, p<0.001$).}
\label{fig_realdata_supp_cka}
\vspace{0em}
\end{figure}

\end{document}